\RequirePackage{ifpdf}
\documentclass[letterpaper]{JHEP3}
\usepackage{amsmath,xspace}
\usepackage[pdftex]{graphicx}
\pdfoutput=1

\newcommand{\Z}{\mathbf{Z}}

\newcommand{\C}{\mathbf{C}}
\newcommand{\CC}{\mathcal{C}}

\newcommand{\OO}{\mathcal{O}}

\DeclareMathOperator{\tr}{tr}

\DeclareMathOperator{\area}{area}

\newcommand{\ev}[1]{\langle{#1}\rangle}
\newcommand{\bev}[1]{\left\langle{#1}\right\rangle}

\title{Entanglement R\'enyi entropies in holographic theories}

\author{
Matthew Headrick \\ 
Martin Fisher School of Physics, Brandeis University, Waltham MA 02453, USA \\ 
\email{mph@brandeis.edu}
}
\abstract{Ryu and Takayanagi conjectured a formula for the entanglement (von Neumann) entropy of an arbitrary spatial region in an arbitrary holographic field theory. The von Neumann entropy is a special case of a more general class of entropies called R\'enyi entropies. Using Euclidean gravity, Fursaev computed the entanglement R\'enyi entropies (EREs) of an arbitrary spatial region in an arbitrary holographic field theory, and thereby derived the RT formula. We point out, however, that his EREs are incorrect, since his putative saddle points do not in fact solve the Einstein equation. We remedy this situation in the case of two-dimensional CFTs, considering regions consisting of one or two intervals. For a single interval, the EREs are known for a general CFT; we reproduce them using gravity. For two intervals, the RT formula predicts a phase transition in the entanglement entropy as a function of their separation, and that the mutual information between the intervals vanishes for separations larger than the phase transition point. By computing EREs using gravity and CFT techniques, we find evidence supporting both predictions. We also find evidence that large-$N$ symmetric-product theories have the same EREs as holographic ones.
}

\preprint{BRX-TH 619}

\begin{document}

\section{Introduction}\label{intro}

The concept of holography originated as an idea about quantum information, that the number of qubits that can be stored in a region of space is fundamentally limited by its surface area in Planck units. Modern holographic theories go beyond a mere counting of states, and posit that the physics governing certain spacetimes can be fully described by a quantum field theory residing on its boundary. However, the way that those qubits are organized remains unclear on both sides of the correspondence. On one side, we don't yet understand how the states are organized in quantum gravity;  on the other, despite an in-principle understanding of the state space of quantum field theories, in practice we have to deal with a strongly coupled theory with a large number of degrees of freedom. And, of course, the map between the two descriptions remains deeply mysterious.

A useful probe of physical information in quantum systems is the entanglement entropy (EE). Here we imagine decomposing a system into two subsystems, $A,A^c$, with a corresponding decomposition of the Hilbert space $\mathcal{H} = \mathcal{H}_A\otimes\mathcal{H}_{A^c}$. Given a density matrix $\rho$ for the full system, the reduced density matrix $\rho_A$, which acts on $\mathcal{H}_A$, is defined by tracing $\rho$ over $\mathcal{H}_{A^c}$ and represents the effective density matrix for an observer who has access only to the subsystem $A$. The EE for $A$ is then the von Neumann entropy of $\rho_A$: $S_A\equiv-\tr(\rho_A\ln\rho_A)$. A non-zero EE may be due to the full system being in a mixed state, to information about the state being lost by the inability to observe the rest of the system, or to a combination of the two effects. The degree of correlation (both classical and quantum) between disjoint subsystems may be quantified by their mutual information $I_{A,B}\equiv S_A+S_B-S_{A\cup B}$, which puts an upper bound on correlators between operators in $A$ and in $B$ \cite{PhysRevLett.100.070502}.

In a quantum field theory, it is natural to consider subsystems that are spatial regions. Their EEs and mutual informations then tell us about the spatial distribution and correlations of quantum information in a given state. Unfortunately, EEs in quantum field theories are notoriously difficult to calculate, mainly because one does not have a good way to represent the operator $\ln\rho_A$. On the other hand, if the density matrix for the full system can be represented by a path integral (as in the vacuum or a thermal ensemble, for example), then both the reduced density matrix $\rho_A$ and its positive integer powers $\rho_A^n$ can also be represented in a fairly simple way by path integrals. If those path integrals can be computed explicitly for all $n$, then one can obtain the EE indirectly as follows. Defining the \emph{entanglement R\'enyi entropy} (ERE) $S_A^{(n)}\equiv(\ln\tr\rho_A^n)/(1-n)$ for $n>1$, one analytically continues $S^{(n)}_A$ in $n$ and takes the limit $n\to1$ to obtain the EE. This procedure is called the replica trick. Aside from being easier to calculate than the EE, the EREs are of interest in their own right, as a more refined characterization of the reduced density matrix $\rho_A$. In fact, knowing $S^{(n)}_A$ for all $n$ is equivalent to knowing the full eigenvalue distribution of $\rho_A$. In Section \ref{Renyireview}, we review the basic properties of entanglement and R\'enyi entropies.

Even given the replica trick, exact results for the EE in field theories are known only in very simple cases, such as a single interval in the vacuum of an arbitrary two-dimensional conformal field theory \cite{Holzhey:1994we}. For two disjoint intervals, the EE, and hence mutual information, remain unknown even for a theory as simple as that of a compact free scalar \cite{Calabrese:2009ez}. It might therefore seem hopeless to dream of knowing the EE in a strongly coupled, large-$N$ field theory. Remarkably, however, Ryu and Takayanagi (RT) proposed a simple, elegant, and universal formula for the EE of an arbitrary spatial region in an arbitrary holographic field theory \cite{Ryu:2006bv,Ryu:2006ef}. Their formula, which applies to any state described by a static classical geometry, says simply that the EE equals one quarter the area in Planck units of the minimal surface in the bulk ending on the boundary of the region $A$. If correct, the RT formula is not only very useful as a calculational tool, but also a significant hint regarding quantum information in holographic theories, and probably in quantum gravity more generally (see for example \cite{VanRaamsdonk:2009ar}).

The RT formula passes several non-trivial checks. For example, it correctly reproduces the EE for a single interval in a two-dimensional CFT. A general derivation, using the replica trick, was offered by Fursaev \cite{Fursaev:2006ih}. He found that the ERE $S_A^{(n)}$ equaled one-quarter the minimal-surface area, independent of $n$. The analytic continuation in $n$ was thus trivial, giving agreement between the resulting value of $S_A$ and the RT formula. In computing the ERE, Fursaev performed the necessary path integrals using Euclidean quantum gravity. Unfortunately, as we show, the bulk geometries that he used to evaluate the partition function are not actually saddle points of the gravitational action. As a result, the ERE he derived is incorrect, as we can see by comparing it to the known exact result in the case of a single interval in a two-dimensional CFT. We show how the latter result can be reproduced using the correct saddle-point action. The RT formula and Fursaev's proof are reviewed and discussed in Section \ref{holographic}.

The question thus arises of whether, in cases where the correct value is not already known, we can compute the ERE in a holographic theory, both for its own sake and in order to confirm or refute the RT conjecture. Unfortunately, to do so in complete generality, as Fursaev attempted, appears to be quite difficult. Therefore, in Section \ref{2disjoint}, we focus on a simple but non-trivial case: two disjoint intervals in a two-dimensional CFT. The RT formula predicts a rather interesting phase transition for the mutual information between the two intervals as a function of their separation. In particular, for separations larger than a certain critical value, the mutual information vanishes, implying a decoupling between the degrees of freedom in the two regions. (This behavior of the mutual information is a completely general prediction of the RT formula, applying essentially to any two regions in any state of any holographic theory. It is closely analogous to the factorization property for disconnected Wilson loops \cite{Gross:1998gk}.)

The ERE for two disjoint intervals can be expressed in terms of the partition function on a certain Riemann surface of genus $n-1$. For $n=2$, we thus need the torus partition function, which fortunately is known for a general holographic CFT \cite{Maldacena:1998bw}. Indeed, as we show, it exhibits a phase transition at precisely the same separation as that predicted for the EE by the RT formula. For higher values of $n$, while the partition function is not known explicitly, we show using symmetry arguments that the ERE continues to have a phase transition at the same separation. This strongly suggests that the same will hold for $n=1$, confirming this prediction of the RT formula.

The fact that we can compute the ERE explicitly only for $n=2$ precludes analytically continuing it to $n=1$, to directly confirm or refute the full EE predicted by the RT formula. We therefore pursue a different strategy. Using the OPE, we expand the ERE, for any given $n$, in powers of the inverse separation between the intervals. The coefficient of any given power can be computed explicitly for all $n$ using formulas for conformal blocks, and analytically continued to $n=1$. We carry this out for a number of coefficients, finding that, thanks to a rather intricate pattern of cancellations, in each case the continuation to $n=1$ vanishes, precisely as predicted by the RT formula.

As a byproduct of our analysis of the ERE for two disjoint intervals, we find that the result for certain non-holographic CFTs with large central charges, such as large-$N$ symmetric-product theories, is precisely the same as for holographic ones. It seems that there is some form of large-$c$ universality operating here, with a large class of such CFTs having identical EREs (and therefore EEs). This possible feature of the ERE deserves further study.

We conclude in Section \ref{discussion} with a list of open questions and possible generalizations of our work, and some remarks concerning our current understanding of the RT formula.

An appendix contains calculations in certain orbifold theories whose results are used in the main text.

\section{Entanglement R\'enyi entropy: review}\label{Renyireview}

In subsection \ref{basic} we briefly motivate, define, and state (without proof) the important properties of the entanglement R\'enyi entropy and mutual R\'enyi information. In subsection \ref{perturbative}, we illustrate these ideas in the simple example of two subsystems that are weakly coupled to each other. In subsection \ref{replica} we then briefly review the replica trick for computing the entanglement R\'enyi entropy, and in subsection \ref{singleinterval} apply it to the simplest field theory example, a single interval in a two-dimensional conformal field theory. For more details, we refer the reader to the books \cite{MR1796805, MR2363070} and the review \cite{calabrese-2009}; the latter provides a comprehensive introduction to R\'enyi and entanglement entropies in two-dimensional CFTs.

\subsection{Basic definitions and properties}\label{basic}

Given a density matrix $\rho$ and a positive real number $\alpha\neq1$, the R\'enyi entropy is defined as\footnote{Writing $e^{-S^{(\alpha)}} = \ev{\rho^{\alpha-1}}_\rho^{1/(\alpha-1)}$, we see that the definition of the R\'enyi entropy is similar to that of the $L^p$ norm of a positive function, $||f||_p \equiv (\int f^p)^{1/p}$. The difference is that, whereas in the $L^p$ norm we evaluate the integral with respect to a fixed measure, in the R\'enyi entropy we evaluate it with respect to the very density matrix whose entropy we are computing.}
\begin{equation}
S^{(\alpha)} \equiv \frac1{1-\alpha}\ln\tr\rho^\alpha\,.
\end{equation}
At $\alpha=1$ the R\'enyi entropy is defined by taking the limit, and equals the von Neumann entropy:
\begin{equation}\label{vonNeumann}
S = S^{(1)} \equiv \lim_{\alpha\to1}S^{(\alpha)} =  -\tr(\rho\ln\rho) \,.
\end{equation}
Two other interesting limits are $\lim_{\alpha\to0}S^{(\alpha)} = \ln\dim\mathcal{H}_{\rm occupied}$, called the Hartley entropy, where $\mathcal{H}_{\rm occupied}$ is the image of $\rho$, and $\lim_{\alpha\to\infty}S^{(\alpha)} = -\ln\rho_{\rm max}$, called the min-entropy, where $\rho_{\rm max}$ is the largest eigenvalue of $\rho$. The following properties of $S^{(\alpha)}$ are straightforward to prove: (1) $S^{(\alpha)}\ge0$, with equality if and only if $\rho$ represents a pure state; (2) $S^{(\alpha)}$ is constant if and only if $\rho$ is proportional to the identity on $\mathcal{H}_{\rm occupied}$, and is otherwise a decreasing function of $\alpha$; (3) for $\alpha>1$ it satisfies $S^{(\alpha)}\le\alpha(\alpha-1)^{-1}S^{(\infty)}$.

If the system contains a subsystem $A$---for example, in a field theory, $A$ could be a spatial region\footnote{By \emph{region} we technically mean co-dimension zero submanifold (possibly with boundary). In this paper we will not consider other types of sets, such as single points, fractals, etc.}---then the Hilbert space $\mathcal{H}$ can be expressed as the tensor product of Hilbert spaces corresponding to $A$ and to its complement $A^{\rm c}$: $\mathcal{H} = \mathcal{H}_A\otimes\mathcal{H}_{A^{\rm c}}$. Let $\rho_A \equiv \tr_{\mathcal{H}_{A^{\rm c}}}\rho$ be the reduced density matrix, defined in $\mathcal{H}_A$, obtained by tracing $\rho$ over  $\mathcal{H}_{A^{\rm c}}$; this is the effective density matrix for an observer who has access only to $A$. Its R\'enyi entropy $S_A^{(\alpha)} = (1-\alpha)^{-1}\ln\tr_A\rho_A^\alpha$ is called the \emph{entanglement R\'enyi entropy} (ERE) of $A$, with the special case $S_A\equiv S_A^{(1)}$ simply called the \emph{entanglement entropy} (EE). It can be shown that, if the full theory is in a pure state, then $S^{(\alpha)}_A = S^{(\alpha)}_{A^{\rm c}}$.

The EE (but not the ERE for $\alpha\neq1$) satisfies an important property called \emph{strong subadditivity} \cite{MR0373508, MR0345558}, namely, for any two subsystems (or spatial regions) $C$ and $D$,
\begin{equation}\label{SSA}
S_C + S_D \ge S_{C\cup D} + S_{C\cap D}\,,\qquad
S_C + S_D \ge S_{C\setminus D} + S_{D\setminus C}\,.
\end{equation}
As a special case, strong subadditivity implies the triangle inequality, namely for disjoint subsystems $A$ and $B$,\footnote{The ERE satisfies $|S_A^{(\alpha)}-S_B^{(\alpha)}|\le S_{A\cup B}^{(\alpha)}$ for classical distributions, but not in general for quantum density matrices.}
\begin{equation}\label{SA}
\left|S_A-S_B\right|\le S_{A\cup B} \le S_A+S_B\,.
\end{equation}
The second inequality is called \emph{subadditivity}, and it characterizes the EE, in the sense that any measure of entanglement that satisfies subadditivity for all subsystems $A$ and $B$ (as well as certain basic requirements such as continuity) must equal the EE \cite{MR0339901,MR0434290}. Subadditivity is saturated if and only if the density matrix $\rho_{A\cup B}$ factorizes: $\rho_{A\cup B} = \rho_A\otimes\rho_B$. Motivated partly by this fact, the \emph{mutual information} (MI) is defined by
\begin{equation}
I_{A,B} = S_A + S_B - S_{A\cup B}\,,
\end{equation}
which quantifies the extent to which the degrees of freedom of $A$ and $B$ are correlated with each other, including both quantum entanglement and classical correlations. For example, the MI puts an upper bound on the connected correlator between (bounded) operators $\mathcal{A}_{A,B}$ in subsystems $A,B$ respectively \cite{PhysRevLett.100.070502}:
\begin{equation}\label{MIinequality}
\left(\frac{\ev{\mathcal{A}_A\mathcal{A}_B}-\ev{\mathcal{A}_A}\ev{\mathcal{A}_B}}
{\|\mathcal{A}_A\|\,\|\mathcal{A}_B\|}
\right)^2
\le 2I_{A,B}\,.
\end{equation}
As a consequence of strong subadditivity, the mutual information is monotone under restriction: if $B'\subset B$ then $I_{A,B'}\le I_{A,B}$.

A natural generalization of the mutual information is to define the \emph{mutual R\`enyi information} (MRI):
\begin{equation}\label{MRIdef}
I^{(\alpha)}_{A,B} \equiv S^{(\alpha)}_A + S^{(\alpha)}_B - S^{(\alpha)}_{A\cup B}\,.
\end{equation}
Unlike the MI, the MRI is not necessarily positive. However, it is non-zero only when $\rho_{A\cup B} \neq \rho_A\otimes\rho_B$, and in this sense still quantifies the extent of correlation between $A$ and $B$.

Another reason to study the MRI (including the MI) is that, when we are considering a field theory, it is universal, whereas the ERE (including the EE) is cutoff- or regulator-dependent. Specifically, when $A$ is a spatial region, $S_A^{(\alpha)}$ usually suffers from an ultraviolet divergence proportional to the area of the boundary of $A$. However, if the two regions $A$ and $B$ are disjoint and mutually disconnected, then those divergences cancel in the MRI. Since the UV regulator generally violates conformal invariance, it follows that in conformal field theories the MRI is generally conformally invariant while the ERE is not. Also, in the CFT case the ERE suffers from an infrared divergence when one of the regions is infinite in size, but this cancels in the MRI (although not when both $A$ and $B$ are infinite). We will see explicit examples of these statements throughout this paper.

\subsection{Perturbative MRI}\label{perturbative}

Before tackling the computation of entanglement entropies in field theories, as a warm-up we first consider the perturbative computation for subsystems that are weakly coupled to each other. We will see that in this case the $\alpha\neq1$ MRI between the subsystems is parametrically larger than the MI, a result that foreshadows the results of Section \ref{2disjoint} concerning holographic systems.

We begin by considering a single system, and the effect on its R\'enyi entropy of a small perturbation to its density matrix:
\begin{equation}
\rho = \rho_{(0)}+\lambda\rho_{(1)}\,,
\end{equation}
where $\tr\rho_{(0)}=1$, $\tr\rho_{(1)}=0$, and $\lambda$ is a small parameter. We assume that the perturbation does not change the rank of the density matrix, and in particular the image of $\rho_{(1)}$ is contained in the image of $\rho_{(0)}$. To first order in $\lambda$ we have:
\begin{align}
S^{(\alpha)} &= \frac1{1-\alpha}\ln\tr\rho_{(0)}^\alpha + \lambda\frac\alpha{1-\alpha}\frac{\tr(\rho_{(1)}\rho_{(0)}^{\alpha-1})}{\tr\rho_{(0)}^{\alpha}} + O(\lambda^2) \\
S &= -\tr(\rho_{(0)}\ln\rho_{(0)})-\lambda\tr(\rho_{(1)}\ln\rho_{(0)}) + O(\lambda^2)\,.
\end{align}

Now suppose our system is composed of two subsystems, and the unperturbed density matrix factorizes:
\begin{equation}
\rho_{(0)} = \hat\rho_A\otimes\hat\rho_B\,.
\end{equation}
At zeroth order in $\lambda$ the MRI of course vanishes. To first order we have:
\begin{equation}\label{MRIperturb}
I_{A,B}^{(\alpha)} = \lambda\frac\alpha{\alpha-1}\tr\left(\rho_{(1)}\left(\frac{\hat\rho_A^{\alpha-1}}{\tr\hat\rho_A^\alpha}-I_A\right)
\otimes
\left(\frac{\hat\rho_B^{\alpha-1}}{\tr\hat\rho_B^\alpha}- I_B\right)
\right) + O(\lambda^2)\,.
\end{equation}
As $\alpha\to1$, the operators $\hat\rho_{A,B}^{\alpha-1}/(\tr\hat\rho_{A,B}^\alpha)-I_{A,B}$ go to zero like $\alpha-1$. Hence, to first order in $\lambda$, the MI vanishes:
\begin{equation}
I_{A,B} = O(\lambda^2)\,.
\end{equation}
(It can be shown that the order $\lambda^2$ term generically does not vanish.) This can be understood as a consequence of the fact that the MI is non-negative, since $\lambda$ could take either sign.

\subsection{Replica trick}\label{replica}

Unfortunately, in practice there are very few known methods for computing EREs (or EEs) in field theories. One of the most useful is the \emph{replica trick}, which we will review below, that allows one to compute the ERE $S^{(n)}$ for integer $n>1$ \cite{Holzhey:1994we}. In favorable circumstances a simple analytic form for $S^{(\alpha)}$ for general real $\alpha$ can be found which fits those data points, and from this form the EE can be read off by setting $\alpha=1$.\footnote{Throughout this paper $\alpha$ will lie in the interval $[0,\infty]$, while $n$ will be a positive integer.} It is important to say at the outset that in proceeding this way we are merely presuming to have guessed the ERE $S^{(\alpha)}$ correctly; firstly, nothing guarantees (in an infinite-dimensional Hilbert space) that the ERE is analytic, and, secondly, the values of a function on a countably infinite set (in this case, the integers larger than 1) are not sufficient to fix a unique analytic continuation. (There exist analytic functions, such as $(1-\alpha)^{-1}\sin\pi\alpha$, that vanish for all integer $\alpha>1$ but not elsewhere, including at $\alpha=1$.) Having stated this caveat, for the rest of the paper we will assume that all EREs we consider are indeed analytic functions of $\alpha$. We will find nothing inconsistent with this assumption.

The replica trick applies when the theory is in a state, such as the vacuum or a thermal state, whose partition function can be obtained by a path integral over some Euclidean spacetime $E$ (possibly with some operator insertions, which for the purposes of this discussion we will consider to be part of $E$). Let $A$ be a spatial region, and $E_n$ the $n$-sheeted cover of $E$ with the sheets connected along branch cuts placed at $A$ on a constant Euclidean-time surface. Then $\tr_A\rho_A^n=Z_n/Z_1^n$, where $Z_n$ is the partition function of the theory on $E_n$ (and, in particular, $Z_1$ is the partition function for the original theory).\footnote{\label{fermionflips}If the theory contains fermions then one needs to specify their boundary conditions across the constant Euclidean-time surface where the sheets are sewn together, which we will call $S$. The original partition function $Z_1$ is computed with a sign-flip on the fermionic fields across $S$. More generally, $Z_n$ is computed with a flip on $S\cap A^{\rm c}$ on each sheet of $E_n$, along with one on $S\cap A$ where the $n$th sheet connects to the first sheet (but not on the other $n-1$ copies of $S\cap A$). Hence a curve that winds around a branch point $n$ times, returning to the same point on $E_n$, crosses $n+1$ sign flips. The resulting overall flip for even $n$ is part of the definition of the twist operators $\sigma_1$ and $\sigma_{-1}$ of the next paragraph. When $Z_n$ is evaluated by passing to a coordinate system that is single-valued on $E_n$, this overall flip is canceled by the branch cut in the coordinate transformation for the fermionic field. (For example, in complex coordinates if $z$ is a local coordinate on $E$ with the branch point at $z=0$, and $t=z^{1/n}$ is a single-valued local coordinate on $E_n$, then $\psi_t = (dz/dt)^{1/2}\psi_z = n^{-1/2}t^{(1-n)/2}\psi_z$. For even $n$ the factor $t^{(n-1)/2}$ has a branch cut with a sign flip.) Hence there is no operator insertion in the new coordinate system. However, there may still be sign flips around non-contractible cycles. (See for example the case of the torus in footnote \ref{torusfermions} below.)} Hence we have (for $n>1$)
\begin{equation}\label{ERE}
S^{(n)}_A = \frac1{1-n}\ln\left(\frac{Z_n}{Z_1^n}\right).
\end{equation}

To be more concrete, let us further specialize to a two-dimensional conformal field theory $\mathcal{C}$,\footnote{All CFTs will be assumed compact, unitary, and modular-invariant in this paper.} and let $A$ be the union of $N$ disjoint intervals $[u_i,v_i]$, where $u_i<v_i<u_{i+1}$. Then we can rewrite the expression \eqref{ERE} in terms of correlators of twist operators in the orbifold theory $\mathcal{C}^n/\Z_n$, computed on $E$:
\begin{equation}\label{ERE2}
S^{(n)}_A = \frac1{1-n}\ln\bev{\sigma^\epsilon_1(u_1)\sigma^\epsilon_{-1}(v_1)\ldots\sigma^\epsilon_1(u_N)\sigma^\epsilon_{-1}(v_N)}\,.
\end{equation}
The correlator of twist operators is divergent, due to the singular geometry of $E_n$ at the branch point. It can be regularized by regularizing each twist operator separately; hence the notation $\sigma^\epsilon_1$ and $\sigma^\epsilon_{-1}$, where $\epsilon$ is a UV cutoff length, for the regularized twist operators \cite{Lunin:2000yv}.

\subsection{Single interval in a CFT}\label{singleinterval}

As an example of the application of \eqref{ERE2}, the ERE for a single interval, in the vacuum, is
\begin{equation}\label{1intervaln}
S^{(n)}_{[u,v]} =\frac1{1-n}\ln\bev{\sigma^\epsilon_1(u)\sigma^\epsilon_{-1}(v)} = \frac c6\left(1+\frac1n\right)\ln\left(\frac{v-u}\epsilon\right) + c_n\,,
\end{equation}
where $c$ is the central charge of $\CC$ and $c_n$ is a scheme-dependent quantity. Here we used the fact that the twist operators have scaling dimension
\begin{equation}\label{twistd}
d_\sigma = \frac c{12}\left(n-\frac1n\right).
\end{equation}
The (simplest) analytic continuation of \eqref{1intervaln} to non-integer $\alpha$ is\footnote{It is interesting to ask what eigenvalue distribution for $\rho_{[u,v]}$ gives rise to the $\alpha$-dependence $S_{[u,v]}^{(\alpha)} = (1+1/\alpha)C$ seen in \eqref{1interval} (neglecting the subleading and scheme-dependent quantity $c_\alpha$). This question can be answered by defining the so-called ``modular Hamiltonian" $\hat H \equiv - \ln\rho_{[u,v]}$ acting on $\mathcal{H}_{[u,v]}$. Then $S^{(\alpha)}_{[u,v]}$ is related to the free energy of $\hat H$ at the temperature $\alpha^{-1}$:
\begin{equation}
F = -\frac1\alpha\ln\tr e^{-\alpha\hat H} = \left(1-\frac1\alpha\right) S^{(\alpha)}_A = \left(1-\frac1{\alpha^2}\right)C\,.
\end{equation}
(The first two equalities apply to the R\'enyi entropy of any system.) The density of states that gives rise to this temperature dependence for the free energy is easily found, in the saddle-point approximation, by performing a Legendre transform:
\begin{equation}\label{DOS}
\rho(\hat E) = \begin{cases} 0\,, & \qquad \hat E<C\\ \exp\left(2C^{1/2}(\hat E-C)^{1/2}\right)\,, & \qquad \hat E>C\end{cases}\,,
\end{equation}
where $\hat E$ is the eigenvalue of $\hat H$. (See \cite{calabrese-2008} or \cite{calabrese-2009} for the form of the full inverse Laplace transform.) It is interesting that (up to a shift of $\hat E$ by $C$) $\rho(\hat E)$ has the same form as the Cardy formula for the asymptotic density of states in a CFT on a circle. Note that we have not determined which physical observable $\hat H$ represents---it is not necessarily related to a physical energy.}
\begin{equation}\label{1interval}
S^{(\alpha)}_{[u,v]} = \frac c6\left(1+\frac1\alpha\right)\ln\left(\frac{v-u}\epsilon\right) + c_\alpha\,,
\end{equation}
which yields the EE \cite{Holzhey:1994we}
\begin{equation}\label{EE1interval}
S_{[u,v]} = \frac c3\ln\left(\frac{v-u}\epsilon\right) + c_1\,.
\end{equation}
Note that $S^{(\alpha)}_{[u,v]}$ indeed satisfies the properties (1), (2), (3) mentioned below equation \eqref{vonNeumann}.

It is also possible to obtain the result \eqref{1intervaln} (and thereby derive the scaling dimension \eqref{twistd}) by computing $Z_n$ and applying \eqref{ERE}. The computation of $Z_n$ is carried out as follows \cite{Lunin:2000yv}. We are in the vacuum, so the Euclidean spacetime $E$ is simply the plane, to which we add a point at infinity to make it topologically a sphere. The multi-sheeted surface $E_n$ is then also topologically a sphere. A Weyl transformation maps the metric $ds^2$ on $E_n$ to a fiducial metric $d\hat s^2 = e^{-\phi}ds^2$ on the sphere. We then have $Z_n = e^{S_L}\hat Z$, where $\hat Z$ is the partition function of $\CC$ on the sphere with the fiducial metric, and $S_L$ is the Liouville action:
\begin{equation}
S_L = \frac c{96\pi}\int \hat g^{1/2}\left(\hat g^{\mu\nu}\partial_\mu\phi\partial_\nu\phi + 2\hat R\phi\right)
\end{equation}
(which depends on $\CC$ only through its central charge). For $n>1$ the metric on $E_n$ has a conical singularity at each branch point $u,v$, so the Liouville action is divergent. The divergence can be regulated by replacing a disc of radius $\epsilon$ about each branch point with a smooth metric, which defines the regularized twist operators $\sigma^\epsilon_{\pm1}$.

\section{Holographic entanglement entropies}\label{holographic}

\subsection{Ryu-Takayanagi formula}\label{RTreview}

In this subsection we will provide a brief summary of Ryu and Takayanagi's proposal for the entanglement entropy (EE) in field theories with holographic duals \cite{Ryu:2006bv,Ryu:2006ef}, along with some of the evidence supporting it. A more complete review can be found in \cite{Nishioka:2009un}. We will then discuss interesting predictions it makes for the mutual information between separated regions.

\subsubsection{Statement}\label{statement}

The Ryu-Takayanagi (RT) conjecture is a proposed formula for the EE of a given spatial region $A$ in certain states of holographic field theories whose dual gravitational theory is classical Einstein gravity (possibly with matter). Specifically, the proposal concerns states that admit a description as static classical solutions in the dual theory, such as the vacuum and thermal states.\footnote{Possible generalizations to time-dependent states were proposed in \cite{Hubeny:2007xt}.} We work in a fixed constant-time (i.e.\ timelike-Killing-field orthogonal) slice of the bulk. The conjecture states that
\begin{equation}\label{RT}
S_A = \frac{\area(m_A)}{4G_{\rm N}}\,,
\end{equation}
where $m_A$ is the minimal-area surface in the bulk that is homologous to $A$, i.e.\ such that there exists a region $r_A$ with $\partial r_A = A\cup m_A$. (As we will see, this topological condition plays a crucial role in several checks of the proposal.) The area is evaluated with respect to the Einstein-frame metric.

An interesting question, assuming the RT formula is valid, is how it gets corrected by quantum effects and by higher-derivative (e.g.\ $\alpha'$) corrections to the classical action in the bulk. Quantum effects presumably lead to $G_{\rm N}$ corrections to \eqref{RT} (starting at order $G_{\rm N}^0$), although a specific form has not been proposed. On the other hand, in the presence of higher-derivative corrections to the classical bulk action, it is expected that the EE is given by minimizing a corrected geometrical functional; based on consistency with black-hole entropy (discussed below), the functional should coincide with Wald's black-hole entropy formula \cite{Wald:1993nt} when evaluated on a horizon.

\subsubsection{Checks}\label{checks}

The RT proposal passes several basic checks. For example, if $A$ is the entire boundary, then $S_A$ should simply be the statistical entropy of the state. Indeed, according to the RT proposal we should take $m_A$ to be the minimal surface in the bulk that is homologous to the boundary; this will generally be the horizon, if there is one, giving agreement with the Bekenstein-Hawking entropy. If there is no horizon, then the boundary is homologically trivial in the bulk (i.e.\ the topological boundary of the bulk is precisely the boundary where the field theory lives); hence the minimal surface is the empty set, giving $S_A = 0$.\footnote{Since we are considering static spacetimes, any black holes in the spacetime should be eternal, so the maximally extended spacetime may include other, topologically disconnected boundaries. Consider, for example, the maximally extended spacetime of the AdS-Schwarzschild black hole, whose boundary has two connected components. It is believed \cite{Maldacena:2001kr} that the field theory defined on \emph{both} boundaries represents the thermofield double of the field theory defined on only one boundary. In this picture, the black hole spacetime, which represents a thermal and therefore mixed state in the original field theory, represents a pure state in the thermofield double. This result is faithfully reproduced by the RT prescription; the full boundary (including both components) is homologically trivial in the bulk, giving $S_A=0$.} (Again, this is the order $G_{\rm N}^{-1}$ entropy---the RT formula does not capture the entropy due for example to a gas of gravitons in thermal AdS, which is of order $G_{\rm N}^0$.) Furthermore, when the total entropy is zero (or of order $G_{\rm N}^0$), then if we instead take $A$ to be a subset of the boundary, we expect from \eqref{SA} that $S_A = S_{A^c}$. Indeed, in this case the entire boundary is homologically trivial in the bulk, so $A$ and $A^c$ are homologous, implying $m_A = m_{A^c}$.

Another important check on the RT proposal is that it satisfies the strong subadditivity (SSA) property \eqref{SSA} for any regions $C$ and $D$, as can be shown by a simple geometrical argument \cite{Headrick:2007km}. (Interestingly, the proof of SSA based on the RT formula is far simpler than the general proof.) Since, as mentioned in subsection \ref{basic}, subadditivity, which is implied by SSA, characterizes the EE, this is quite strong evidence in favor of the RT formula. However, it is not sufficient to prove its correctness, since it only shows that \eqref{SSA} is satisfied for subsystems corresponding to geometrical regions, whereas for the characterization proof one needs it to hold for \emph{all} subsystems. The proof of SSA extends trivially to the inclusion of higher-derivative corrections, as long as they are extensive.

As a final check, let us see how the RT formula reproduces the EE \eqref{EE1interval} of a single interval $[u,v]$, in the vacuum of a two-dimensional CFT. The vacuum is described holographically by AdS${}_3$, whose metric on a constant-time slice is
\begin{equation}
ds^2 = \frac{\ell_{\rm AdS}^2}{z^2}\left(dz^2+dy^2\right);
\end{equation}
here $y$ is the coordinate along the boundary and $z$ is the radial coordinate, with the boundary being at $z=0$. We employ a simple UV cutoff in which we shift the boundary to $z=\epsilon$. The minimal surface $m_{[u,v]}$ is a geodesic connecting the points on the boundary $(y,z) = (u,\epsilon),(v,\epsilon)$, which is an arc of a circle (almost a semi-circle) with center $((u+v)/2,0)$. Applying \eqref{RT} and using the standard holographic relation $\ell_{\rm AdS}/G_{\rm N} = 2c/3$, one finds \cite{Ryu:2006bv,Ryu:2006ef}
\begin{equation}\label{RT1interval}
S_{[u,v]} = \frac{\ell_{\rm AdS}}{2G_{\rm N}}\ln\left(\frac{v-u}\epsilon\right)
 = \frac c3\ln\left(\frac{v-u}\epsilon\right),
\end{equation}
matching \eqref{EE1interval}. (In this scheme, the finite part $c_1$ vanishes.) In higher dimensional CFTs, although one does not have exact formulas for the EEs even of simple regions, the leading UV divergence is known and matches that predicted by the RT formula \cite{Ryu:2006bv,Ryu:2006ef}.

\subsubsection{Application to disconnected regions}\label{RTdisconnected}

For a time it was believed that the RT formula should only apply to connected regions. (See for example the paper \cite{Hubeny:2007re}.) The reason was that, when applied to the union of two intervals (a case that will be considered in detail in the next section), it disagreed with a calculation by Calabrese and Cardy \cite{Calabrese:2004eu} which (like the formula \eqref{EE1interval} for a single interval) was supposed to be valid in any two-dimensional CFT. However, those same authors have since shown that their original calculation was incorrect. At present, there is no reason to believe that the RT formula, if it is valid at all, would not apply equally well to connected and to disconnected regions. For example, all the checks discussed above apply to both cases (including the last one, which can be considered a computation of the EE of the disconnected region $(-\infty,u]\cup[v,\infty)$).

When applied to a disconnected region, the RT formula makes a fascinating prediction for the mutual information (MI) between its components, similar to the phase transition for disconnected Wilson loops found by Gross and Ooguri \cite{Gross:1998gk}. For simplicity, let us consider two disjoint and mutually disconnected regions $A$, $B$. Each has a corresponding minimal surface $m_A$, $m_B$ and region $r_A$, $r_B$. (We assume the generic situation that $r_A$, $r_B$ are disjoint and mutually disconnected.) When we consider the region $A\cup B$, the disconnected surface $m_A\cup m_B$ is topologically allowed and locally minimal. Assuming that the full bulk spacetime is itself connected, surfaces will also exist that connect $A$ and $B$. However, if the separation between $A$ and $B$ is sufficiently large compared to their sizes (and any other scales defined in the theory or state), then $m_A\cup m_B$ will necessarily be the globally minimal surface. Then we have $S_{A\cup B}=S_A+S_B$, so the MI $I_{A,B}$ vanishes. More precisely, $I_{A,B}$ is of order $G_{\rm N}^0$, rather than $G_{\rm N}^{-1}$.\footnote{A closely related phenomenon, in which the $A$ and $B$ are held fixed but the bulk spacetime is deformed, was discussed in the paper \cite{VanRaamsdonk:2009ar}.} This implies that, from a quantum information point of view, the two regions are approximately decoupled from each other. (See for example the bound \eqref{MIinequality} on correlators between $A$ and $B$. Note however that this bound does not directly give us information about correlators of local operators, which are generally not bounded operators.) If we then imagine bringing $A$ and $B$ closer to each other, then it may happen that, at some critical separation, the minimal surface will switch from $m_A\cup m_B$ to one that connects $\partial A$ and $\partial B$ (see for example figure \ref{minimal}). In this case, the MI will (in the thermodynamic/classical limit $G_{\rm N}\to0$) undergo a first-order phase transition; it will become non-zero, with a continuous value but discontinuous first derivative as a function of the separation between $A$ and $B$. Section \ref{2disjoint} will be devoted to a detailed study of these phenomena in the simplest example, namely two intervals in the vacuum of a two-dimensional CFT.

\subsection{Fursaev's ERE calculation}\label{Fursaevsubsection}

In the paper \cite{Fursaev:2006ih}, Fursaev gave a derivation, based on the replica trick, of the RT formula. In this subsection, we will briefly summarize his argument, and then point out a flaw that results in an incorrect value for the entanglement R\'enyi entropy (ERE).

In our sketch of Fursaev's argument, for simplicity we will take the bulk action to be pure Einstein gravity; matter fields and higher-derivative (e.g.\ $\alpha'$) corrections are straightforwardly incorporated, as he discusses. We will also assume that the ultraviolet divergence in the field theory is cut off in some manner whose details will not concern us. Fursaev's starting point is \eqref{ERE}, where $Z_n$ is the partition function on the $n$-sheeted Euclidean spacetime $E_n$. Recall that, if $A$ is the spatial region whose EE we are computing, then the sheets of $E_n$ are connected by a branch cut along $A$ on a constant-time slice. In a holographic theory, this partition function is given by the gravitational path integral over Euclidean geometries whose conformal boundary is $E_n$. In the classical limit, this path integral goes over to its saddle-point approximation $e^{-S_{\rm min}}$, where $S_{\rm min}$ is the minimal value of the Euclidean Einstein-Hilbert action among extrema obeying the boundary conditions. Fursaev constructs a set of geometries with boundary $E_n$, then minimizes the Euclidean action within that set. He takes as given the bulk Euclidean spacetime $\tilde E$ representing the original state of the system; its boundary is $E$ and its Euclidean action is $-\ln Z_1$. He takes $n$ copies of $\tilde E$ and connects them along a branch cut $r_A$, which is a spatial region in $\tilde E$ lying in the same constant-time slice as $A$. In order for this $n$-sheeted bulk geometry to have boundary $E_n$, the part of the boundary of $r_A$ that lies in $E$ must coincide with $A$ (i.e.\ $\partial r_A\cap E = A$); apart from this condition, the choice of $r_A$ is at this point arbitrary. The branch ``point" is $m_A$, the rest of the boundary of $r_A$ ($m_A = \partial r_A\setminus A$ and $\partial r_A = A\cup m_A$). He now evaluates the Euclidean Einstein-Hilbert action for this geometry. There are two contributions. First, the geometry is made up of $n$ copies of $\tilde E$, so there is a contribution $-n\ln Z_1$, which is independent of the choice of $r_A$. In addition, the Ricci scalar has a delta function along the branch ``point" $m_A$, which is co-dimension 2 and hosts a conical singularity with excess angle $2\pi(n-1)$. It therefore contributes a term $(n-1)\area(m_A)/(4G_{\rm N})$ to the action. Minimizing this action over all possible choices of $r_A$, he obtains the minimal surface $m_A$, and (from \eqref{ERE}) the ERE
\begin{equation}\label{Fursaev}
S_A^{(n)} = \frac{\area(m_A)}{4G_{\rm N}}\,.
\end{equation}
Since there is no $n$-dependence, the analytic continuation is particularly simple: $S^{(\alpha)}_A = \area(m_A)/4G_{\rm N}$. Finally, setting $\alpha=1$, he obtains the RT formula \eqref{RT}.

The problem with this derivation is that the action has been extremized only with respect to a subset of the degrees of freedom in the metric, namely the choice of $r_A$. The resulting field configuration is therefore not guaranteed to be a true saddle-point, and in fact it does not solve the Einstein equation: the Einstein tensor has a delta function supported on $m_A$ due to the conical singularity, with no corresponding source.

We can confirm that the ERE \eqref{Fursaev} is incorrect by comparing it to the exact result in a case where the latter is known. For example, when $A$ is a single interval in the vacuum of a two-dimensional CFT, the exact ERE \eqref{1intervaln} depends on $n$ (by the factor $1+1/n$), whereas the Fursaev result \eqref{Fursaev} is independent of $n$. What is the true saddle point in this case? The Euclidean space $E$ is a plane, and the corresponding bulk geometry $\tilde E$ is hyperbolic 3-space $H^3$ (a.k.a.\ Euclidean AdS${}_3$). The saddle point corresponding to the $n$-sheeted cover $E_n$ is also $H^3$. The easiest way to see this is to add a point at infinity to $E$ to make it a sphere; then its $n$-sheeted cover $E_n$ is also a sphere, so the corresponding bulk geometry is $H^3$. This geometry is smooth, in contrast to Fursaev's, which is $n$ copies of $H^3$ glued together in such a way as to create a conical singularity along the geodesic connecting the endpoints of $A$. Given that the bulk geometry is $H^3$ for all $n$, why does its action depend on $n$? The bulk action is divergent due to the infinite volume near the boundary; while the full bulk geometry is $H^3$ for any $n$, the cutoff geometry depends on $n$. (The full geometry depends only on the Weyl class of the boundary metric, which is the same for all $n$, since the sphere admits a unique Weyl class. On the other hand, the cutoff geometry is sensitive to the actual boundary metric. This is the holographic manifestation of the Weyl anomaly \cite{Henningson:1998gx}.) The $n$-dependence can most easily be calculated by performing a Weyl transformation to put the metric on $E_n$ into a standard form and taking into the account the resulting change in the partition function due to the Liouville action, as described at the end of subsection \ref{singleinterval}, or equivalently by the holographic renormalization procedure \cite{Henningson:1998gx}.

It is worth noting that the true $H^3$ saddle point with boundary $E_n$ can be obtained \emph{topologically} by gluing $n$ copies of $\tilde E = H^3$ together in precisely the manner described by Fursaev. This will continue to be the case in the more complicated examples we will study in the next section, suggesting that, while it carries the wrong metric, Fursaev's construction may be topologically correct in general. This would explain why the topological condition on the minimal surface $m_A$ that he suggested---that $m_A$ should be homologous to $A$---appears to be correct.

Finally, it is intriguing that, while Fursaev's value \eqref{Fursaev} for the ERE is incorrect for $n>1$, it somehow manages to give the right answer for the EE ($n=1$), assuming that the RT conjecture holds. We can only speculate that, if there is some sense in which the spacetimes $E_n$ and their bulk duals can be defined for non-integer values of $n$, then his construction may be correct ``at linear order" in a neighborhood of $n=1$.

\section{Mutual R\'enyi information between two intervals}\label{2disjoint}

As we saw in subsection \ref{singleinterval}, the entanglement entropy for a single interval in the vacuum of a two-dimensional CFT depends only on the theory's central charge. The fact that the Ryu-Takayanagi formula correctly reproduces this entropy, as reviewed in subsection \ref{statement}, is an important check on the proposal, but does not give us any new information. The next simplest configuration we can consider in such a theory consists of two disjoint intervals. As suggested by the fact that the R\'enyi entropies \eqref{ERE2} depend in this case on four-point rather than two-point functions of twist operators, we would expect the EE to depend on the full operator content of the theory, rather than simply its central charge. As we will see in this section, the RT formula can give us significant new physical information in this case. The new predictions in turn give us the opportunity to subject the formula to new and highly non-trivial quantitative tests.

We begin by reviewing the necessary formulas and setting up the basic properties of the ERE for two intervals.

\subsection{General properties}\label{2disjointgeneral}

We consider two separated intervals $[u_1,v_1]$, $[u_2,v_2]$ ($u_1<v_1<u_2<v_2$) in the vacuum of a conformal field theory $\mathcal{C}$ with central charge $c$. As discussed in subsection \ref{basic}, it is convenient to consider the mutual R\'enyi information (MRI) between the two intervals,
\begin{equation}\label{MRIdef2}
I^{(\alpha)}_{[u_1,v_1],[u_2,v_2]} = 
S^{(\alpha)}_{[u_1,v_1]} + S^{(\alpha)}_{[u_2,v_2]} -S^{(\alpha)}_{[u_1,v_1]\cup[u_2,v_2]}\,,
\end{equation}
which measures the extent to which the degrees of freedom of the two intervals are entangled with each other (including both classical correlations and quantum entanglement).

We first consider the integer case $\alpha = n > 1$. Using \eqref{ERE2}, the MRI is given in terms of a finite ratio of four-point and two-point functions of twist operators in the orbifold theory $\CC^n/\Z_n$:
\begin{align}
I^{(n)}_{[u_1,v_1],[u_2,v_2]}
&= \frac1{n-1}\ln\left(\frac
{\ev{\sigma^\epsilon_1(u_1)\sigma^\epsilon_{-1}(v_1)\sigma^\epsilon_1(u_2)\sigma^\epsilon_{-1}(v_2)}}
{\ev{\sigma^\epsilon_1(u_1)\sigma^\epsilon_{-1}(v_1)}\ev{\sigma^\epsilon_1(u_2)\sigma^\epsilon_{-1}(v_2)}}
\right) \\
&= \frac1{n-1}\ln\left(\frac
{\ev{\sigma_1(u_1)\sigma_{-1}(v_1)\sigma_1(u_2)\sigma_{-1}(v_2)}}
{\ev{\sigma_1(u_1)\sigma_{-1}(v_1)}\ev{\sigma_1(u_2)\sigma_{-1}(v_2)}}
\right),
\end{align}
where we've defined the renormalized twist operators:
\begin{equation}
\sigma_{\pm1} \equiv\frac {\sigma^\epsilon_{\pm1}}{\ev{\sigma^\epsilon_1(0)\sigma^\epsilon_{-1}(1)}^{1/2}}\,.
\end{equation}
This is an example of the UV divergences in the EREs, which occur at the endpoints of the intervals, cancelling in the MRI, as discussed at the end of subsection \ref{basic}.

Since the twist operators are primaries, the transformation law for the four- and two-point functions implies that $I_{[u_1,v_1],[u_2,v_2]}$ is conformally invariant, and therefore depends only on the cross-ratio
\begin{equation}\label{crossdef}
x \equiv \frac{(v_1-u_1)(v_2-u_2)}{(u_2-u_1)(v_2-v_1)}\,,
\end{equation}
which lies in the interval $0<x<1$. By a conformal transformation, the four points $u_1$, $v_1$, $u_2$, $v_2$ can be brought to 0, $x$, 1, $\infty$ respectively, so we have:
\begin{equation}\label{MRI2intervals}
I^{(n)}_{[u_1,v_1],[u_2,v_2]}
= I^{(n)}(x)
\equiv I^{(n)}_{[0,x],[1,\infty]}
= \frac1{n-1}\ln\left(x^{2d_\sigma}\bev{\sigma_1(0)\sigma_{-1}(x)\sigma_1(1)\sigma_{-1}'(\infty)}\right),
\end{equation}
where $\sigma'_{-1}(\infty)\equiv\lim_{z\to\infty}z^{2d_\sigma}\sigma_{-1}(z)$. (The scaling dimensions $d_\sigma$ of the twist operators are given by \eqref{twistd}.) Notice that, like the UV divergence, the IR divergence in the ERE cancels in the MRI. The four-point function, and therefore $I^{(n)}(x)$, is an analytic function of $x$ in the interval $0<x<1$.

It is useful to note that the four-point function in \eqref{MRI2intervals} can be expanded as a power series in $x$, where the powers are the dimensions $d_m$ of operators $\mathcal{A}'_m$ in the orbifold theory,\footnote{Throughout the paper, we use primes on untwisted operators of $\CC^n/\Z_n$, to distinguish them from operators of $\CC$.} and the coefficients are given in terms of OPE coefficients:
\begin{equation}\label{MRIOPE}
I^{(n)}(x)
= \frac1{n-1}\ln\left(\sum_m{c^{\sigma_1}}_{\sigma_1m}{c^m}_{\sigma_1\sigma_{-1}}x^{d_m}\right).
\end{equation}
Note that only untwisted operators contribute to the sum. Assuming we are dealing with a unitary theory, the operator with lowest scaling dimension is the unit operator, for which (by the normalization of the twist operators) the OPE coefficients are 1. Hence $I^{(n)}(x)$ goes to 0 as $x\to0$, as we would expect on physical grounds. For example, if we fix the sizes of the intervals and take their separation to infinity, we would expect all correlations between them to go to zero. We will study the higher-order terms in the expansion \eqref{MRIOPE} in subsections \ref{phasetransition} and \ref{expansionx}, and in the appendix.

A final important property of the MRI, implied by the invariance of the four-point function in \eqref{MRI2intervals} under $x\to1-x$, is
\begin{equation}
I^{(n)}(1-x) = I^{(n)}(x) + \frac c6\left(1+\frac1n\right)\ln\frac{1-x}x\,.
\end{equation}
At the level of the definition \eqref{MRIdef2} of the MRI, this relation is due to the fact that, in a pure state (in this case, the vacuum), $S^{(\alpha)}_A = S^{(\alpha)}_{A^c}$, so $S^{(\alpha)}_{[0,x]\cup[1,\infty]} = S^{(\alpha)}_{[-\infty,0]\cup[x,1]} = S^{(\alpha)}_{[0,1-x]\cup[1,\infty]}$.

We have listed five general properties that the MRI satisfies for integer $\alpha>1$, but for the reasons given we either know or expect each to hold for general values of $\alpha$:
\begin{enumerate}
\item UV finiteness, and IR finiteness when one of the intervals is semi-infinite;
\item conformal invariance, implying
\begin{equation}
I^{(\alpha)}_{[u_1,v_1],[u_2,v_2]} = I^{(\alpha)}(x) \equiv I^{(\alpha)}_{[0,x],[1,\infty]}
\end{equation}
(where $0<x<1$);
\item\begin{equation}\label{limxto0}
\lim_{x\to0}I^{(\alpha)}(x) = 0\,;
\end{equation}
\item for all $x$,
\begin{equation}\label{x1-x}
I^{(\alpha)}(1-x) = I^{(\alpha)}(x) + \frac c6\left(1+\frac1\alpha\right)\ln\frac{1-x}x\,;
\end{equation}
\item analyticity of $I^{(\alpha)}(x)$ as a function of $x$.
\end{enumerate}
  
So far we have not assumed anything about the theory $\CC$ (other than unitary and compactness). In the rest of this section, we will study the function $I^{(\alpha)}(x)$ in holographic CFTs, as well as certain other theories with large central charge.

\subsection{Prediction from Ryu-Takayanagi formula}\label{RTprediction}

As in the holographic derivation of the EE for a single interval, reviewed in subsection \ref{checks}, we use the fact that the holographic dual of the vacuum is AdS${}_3$, with $\ell_{\rm AdS}/G_{\rm N} = 2c/3$, and we cut off integrals near the boundary at radial coordinate value $z=\epsilon$.

\FIGURE{
\includegraphics[width=6.1in]{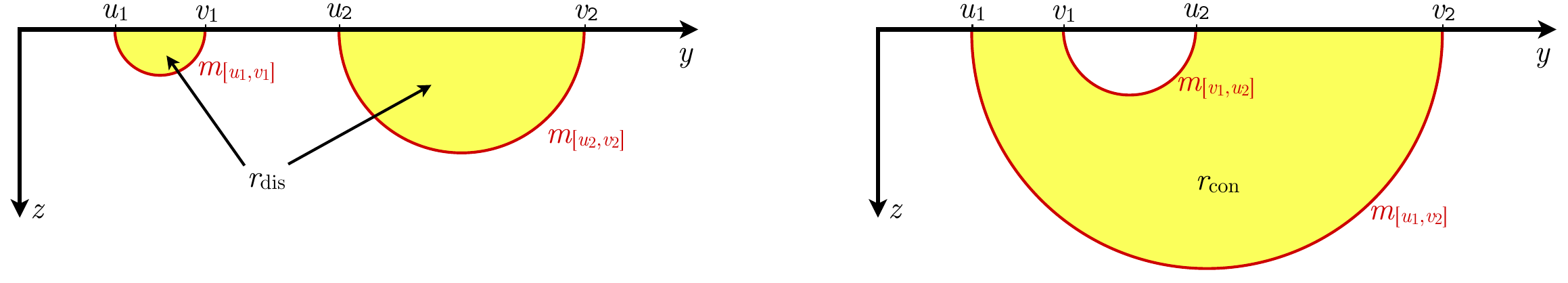}
\caption{The two locally minimal surfaces for the boundary region $[u_1,v_1]\cup[u_2,v_2]$. The global minimum is the one on the left is when $x<1/2$, and the one on right when $x>1/2$, where $x$ is the cross-ratio defined in \eqref{crossdef}.}
\label{minimal}
}

The RT formula is straightforward to apply to the union of two intervals $[u_1,v_1]\cup[u_2,v_2]$. There are two locally minimal surfaces in the bulk that are homologous to this boundary region, as shown in figure \ref{minimal}. The first is the union of the minimal surfaces for the two intervals separately,  $m_{\rm dis} = m_{[u_1,v_1]}\cup m_{[u_2,v_2]}$ (similarly for the corresponding bulk region $r_{\rm dis} = r_{[u_1,v_1]}\cup r_{[u_2,v_2]}$). This has ``area" (i.e.\ length)
\begin{align}\label{disconnected}
\area(m_{\rm dis}) &=
\area(m_{[u_1,v_1]}) + \area(m_{[u_2,v_2]})  \\
&= 2\ell_{\rm AdS}\ln\left(\frac{v_1-u_1}\epsilon\right) + 2\ell_{\rm AdS}\ln\left(\frac{v_2-u_2}\epsilon\right)
\end{align}
(see \eqref{RT1interval}). The other locally minimal surface connects $u_1$ to $v_2$ and $u_2$ to $v_1$: $m_{\rm con} = m_{[u_1,v_2]}\cup m_{[v_1,u_2]}$. (The corresponding bulk region is a semi-annulus connecting the two intervals: $r_{\rm con} = r_{[u_1,v_2]}\setminus r_{[v_1,u_2]}$.) Its area is
\begin{equation}\label{connected}
\area(m_{\rm con}) =
2\ell_{\rm AdS}\ln\left(\frac{v_2-u_1}\epsilon\right) + 2\ell_{\rm AdS}\ln\left(\frac{u_2-v_1}\epsilon\right).
\end{equation}
It is easy to see that $m_{\rm dis}$ is the globally minimal surface when $x<1/2$, and $m_{\rm con}$ otherwise (see \eqref{crossdef}), so
\begin{align}\label{EE2intervals}
S_{[u_1,v_1]\cup[u_2,v_2]} &= \frac1{4G_{\rm N}}\min(\area(m_{\rm dis}),\area(m_{\rm con})) \nonumber \\ &= 
\frac c3\times\begin{cases}
\ln((v_1-u_1)(v_2-u_2)/\epsilon^2)\,,\quad & x\le1/2 \\
\ln((v_2-u_1)(u_2-v_1)/\epsilon^2)\,,\quad & x\ge1/2
\end{cases}\,.
\end{align}

Combining \eqref{EE2intervals} with \eqref{RT1interval}, we obtain the following mutual information:
\begin{equation}\label{RTMI}
I_{[u_1,v_1],[u_2,v_2]} = I^{(1)}(x) = \begin{cases} 0\,,\quad & x\le1/2 \\ (c/3)\ln(x/(1-x))\,,\quad & x\ge1/2\end{cases}\,.
\end{equation}
Of the five properties of the MI listed at the end of the previous subsection, this formula obeys the first four. It does not obey the last---analyticity---as it has a discontinuous first derivative at $x=1/2$ and vanishes for $x\le1/2$. These two features were anticipated in the discussion in subsection \ref{RTdisconnected}. The discontinuity in the first derivative occurs because the global minimum switches between the local minima as we vary $x$, and is reminiscent of phase transitions due to competing saddle points of the Euclidean action, such as the Hawking-Page transition. As in that case, the transition is presumably sharp only in the classical limit in the bulk, which corresponds to the thermodynamic ($c\to\infty$) limit of the CFT, and gets smoothed out by finite-$c$ effects. Similarly, the vanishing of the MI for $x\le1/2$ is presumably true only at order $c$; if the MI vanished exactly for $x\le1/2$, then the reduced density matrix for the two intervals would factorize, $\rho_{[u_1,v_1]\cup[u_2,v_2]} = \rho_{[u_1,v_1]}\otimes\rho_{[u_2,v_2]}$, implying that the two intervals are completely decoupled from each other; in particular, it would imply that all connected correlators vanish, which is certainly not the case. Thus we should expect both perturbative and non-perturbative corrections to \eqref{RTMI} in $G_{\rm N}\sim c^{-1}$, with the first perturbative correction at order $c^0$. Nonetheless, since the MI is apparently parametrically small for $x\le1/2$---smaller than the EE for either interval separately or for their union, and smaller than the MI for $x>1/2$---it appears that the density matrix factorizes approximately.

Unlike quantum corrections, we do not expect higher-derivative (e.g.\ $\alpha'$) corrections to the classical bulk action to change the result \eqref{RTMI}, for the following reason. As discussed in subsection \ref{statement}, such corrections are believed to correct the area functional appearing in the RT formula without changing the basic prescription of minimizing over topologically allowed surfaces. The symmetries of AdS${}_3$ guarantee that the minimal surfaces shown in figure \ref{minimal} remain uncorrected; furthermore, the corrected ``area" of each curve is unchanged when written as a function of $c$, since we know that the EE is always given by \eqref{EE1interval}. In fact, this argument applies for any bulk gravitational action, not just Einstein-Hilbert with small higher-derivative corrections.

\subsection{Universality in the large-$c$ limit?}\label{universality}

Given any family of CFTs $\CC$ that admit a large-$c$ limit, such as holographic ones, we can consider the expansion of the MRI in powers of $c^{-1}$. Since the number of degrees of freedom is of order $c$, the leading term will be at most of that order, so we have
\begin{equation}
I^{(\alpha)}(x) = I_1^{(\alpha)}(x)c + I_0^{(\alpha)}(x) + O(c^{-1})\,.
\end{equation}
In particular, we focus our attention on the leading function $I_1^{(\alpha)}(x)$. In the previous subsection, we used the RT formula to compute, for holographic theories,
\begin{equation}\label{I11}
I_1^{(1)}(x) =  \begin{cases} 0\,,\quad & x\le1/2 \\ (1/3)\ln(x/(1-x))\,,\quad & x\ge1/2\end{cases}\,,
\end{equation}
and argued that this result should hold no matter what the bulk gravitational theory is. As discussed, \eqref{I11} has two striking qualitative features, namely its discontinuous first derivative at $x=1/2$ and the fact that it vanishes for $x\le1/2$. In the rest of this section, we will study $I_1^{(\alpha)}(x)$ using the replica trick, and find independent evidence for both phenomena. In the next subsection, we will compute $I_1^{(2)}(x)$ in holographic theories, and show that the result is independent of the details of the bulk theory (e.g.\ the presence of higher-curvature corrections), and applies also to symmetric-product theories $\CC = \CC_0^N/S_N$, even though their large-$c$ limit is not described by classical gravity. Like \eqref{I11}, the result will have a phase transition at $x=1/2$. In subsection \ref{MRIng2} we will argue that this phase transition occurs also in $I_1^{(n)}(x)$ for $n>2$, at least in holographic theories. Then in the last subsection we will use CFT techniques to study the expansion of $I_1^{(\alpha)}(x)$ in powers of $x$ for general $\alpha$, and find evidence that every coefficient in this expansion goes to 0 in the limit $\alpha\to1$. That analysis will assume very little about the CFT $\CC$, essentially just that the number of operators below any given dimension stays finite as $c$ goes to infinity, a condition that holds for both holographic and symmetric-product theories (but not, for example, in the power theory $\CC_0^N$ without the orbifold).

These results not only give strong quantitative support to the RT formula, but point to a broader picture, namely that a large class of large-$c$ CFTs---including holographic and symmetric-product theories---share the same leading MRI $I^{(\alpha)}_1(x)$, as a function of both $\alpha$ and $x$. Although we do not know the explicit form of this function except for $\alpha=1,2$, we can deduce that it is analytic in $x$ except at $x=1/2$, where it has a discontinuous first derivative, and satisfies the following properties:
\begin{align}
\lim_{x\to0}I_1^{(\alpha)}(x) &= 0\,,\\
I_1^{(\alpha)}(1-x) &= I_1^{(\alpha)}(x) + \frac16\left(1+\frac1\alpha\right)\ln\frac{1-x}x\,.
\end{align}

Based on these considerations, it appears that in the range $0<x\le1/2$ the MRI is parametrically larger for $\alpha\neq1$ (where it is of order $c$) than for $\alpha=1$ (where it is of order $1$). This is similar to what we found in the perturbative calculation of subsection \ref{perturbative}. It would be interesting to find a simple toy model of a system with $N$ degrees of freedom, in which the MRI between two subsystems is of order $N$, but the MI is only of order $1$.

\subsection{MRI for $n=2$}\label{phasetransition}

In this subsection we will begin by expressing the mutual R\'enyi information $I^{(2)}(x)$ in a general CFT in terms of its torus partition function. Using this expression, we will calculate the order-$c$ part $I^{(2)}_1(x)$ in a general holographic CFT, finding that---like the RT prediction \eqref{I11} for $I_1^{(1)}(x)$---it is analytic except at $x=1/2$, where it has a discontinuous first derivative. We will then show that $I^{(2)}_1(x)$ is precisely the same function in large-$N$ symmetric-product theories, supporting the idea of universality (i.e.\ theory-independence in the large-$c$ limit) proposed in the previous subsection.

\FIGURE{
\includegraphics[width=2.2in]{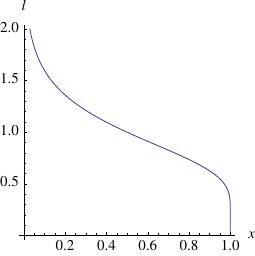}
\caption{Modular parameter $\tau = il$ for the two-sheeted Riemann surface with branch points at $0$, $x$, $1$, $\infty$. The relation between $x$ and $l$ is given by equation \eqref{xtau}.}
\label{lvsx}
}

\subsubsection{General CFTs}

We begin by applying the formula \eqref{MRI2intervals} for $n=2$. In the $\CC^2/Z_2$ orbifold theory, there is a unique twist operator $\sigma\equiv\sigma_1 = \sigma_{-1}$. Lunin and Mathur \cite{Lunin:2000yv} showed that its four-point function is given by\footnote{In terms of Lunin and Mathur's variables, $x=1/w$ and $il = \tau = -1/\tau_{\text{Lunin-Mathur}}$. The four-point function of twist fields was computed in \cite{Dixon:1986qv} in the case where the underlying CFT is a free scalar field.}
\begin{equation}\label{4point}
\ev{\sigma(0)\sigma(x)\sigma(1)\sigma'(\infty)}
= \left(2^8x(1-x)\right)^{-c/12}Z_{il}\,,
\end{equation}
where $Z_{il}$ is the partition function for $\mathcal{C}$ on a flat rectangular torus\footnote{\label{torusfermions}The fermion sign flips explained in footnote \ref{fermionflips} imply that fermionic fields should have antiperiodic (NS) boundary conditions on both cycles of the torus. The reason is that, on the double-sheeted plane, in going around either cycle one encounters two sign flips, so no overall flip. The coordinate transformation to the torus introduces a flip, just as when passing from the plane to the cylinder. Hence the partition function is not invariant under the full modular group, but it is invariant under $\tau\to-1/\tau$.} with modular parameter $\tau = il$; $x$ and $l$ are related by
\begin{equation}\label{xtau}
x = \frac{\theta_2^4(il)}{\theta_3^4(il)}\,.
\end{equation}
As $x$ goes from 0 to 1, $l$ goes from $\infty$ to 0, with $x=1/2$ corresponding to $l=1$ (see figure \ref{lvsx}). Since
\begin{equation}\label{1-x}
1-x = \frac{\theta_4^4(il)}{\theta_3^4(il)} =  \frac{\theta_2^4(i/l)}{\theta_3^4(i/l)}\,,
\end{equation}
the invariance of the four-point function \eqref{4point}, and hence of the ERE, can be traced to the modular invariance of the torus partition function, $Z_{il} = Z_{i/l}$. The reason for the appearance of the torus partition function of $\CC$ is that the four-point function of twist operators is the (renormalized) zero-point function on the two-sheeted Riemann surface $E_2$ with a branch cut connecting $0$ to $x$ and another one connecting $1$ to $\infty$, which is a torus with complex structure $\tau=il$. The Weyl transformation that flattens it, when plugged into the Liouville action, leads to the prefactor $(2^8x(1-x))^{-c/12}$.

Plugging \eqref{4point} into \eqref{MRI2intervals}, and using the identity $2\eta^3=\theta_2\theta_3\theta_4$, we obtain
\begin{align}\label{MCI}
I^{(2)}(x)
 &= \ln Z_{il} - \frac c{12}\ln\left(\frac{2^8(1-x)}{x^2}\right)\nonumber \\ &=\ln  Z_{il} + c\ln\left(\frac{\theta_2(il)}{2\eta(il)}\right).
\end{align}
The first term in \eqref{MCI} is ($-l$ times) the free energy of $\CC$ on a circle of unit circumference at temperature $l^{-1}$. (Note however that the basic cycles of this torus, which we interpret as space and Euclidean time directions when we speak of the free energy, are not the same as the space and time directions of the Euclidean plane $E$ where the theory was originally defined, whose double cover is $E_2$. Rather, the spatial circle of the torus encircles the points $0$ and $x$, staying on one sheet, while its Euclidean time circle encircles $x$ and $1$, crossing each branch cut once.)

Let us consider the expansion of \eqref{MCI} for small values of $x$, where $x\approx 16e^{-\pi l}$.\footnote{Defining $q = e^{2\pi i\tau} = e^{-2\pi l}$, the expansions of the theta and eta functions for small $q$ are as follows: $\theta_2(\tau) = 2q^{1/8}(1+q+O(q^3))$, $\eta(\tau) = q^{1/24}(1-q+O(q^2)).$ Hence $\theta_2(\tau)/2\eta(\tau) = q^{1/12}(1+2q+O(q^2)).$} The behavior of the torus partition function is universal in this limit (for compact unitary CFTs), $\ln Z_{il}\approx 2\pi cl/12$, which precisely cancels the leading behavior of the second term in \eqref{MCI}, giving a vanishing MRI as expected (equation \eqref{limxto0}). The leading $x$-dependence depends on the gap in the operator spectrum of $\mathcal{C}$. If the lowest non-unit operator $\hat{\mathcal{A}}$ has dimension $\hat d$ and multiplicity $\hat m$ (where fermionic operators are counted negatively), then
\begin{equation}
\ln Z_{il} = \frac{2\pi cl}{12} + \hat me^{-2\pi\hat dl} + \cdots\,,
\end{equation}
so
\begin{equation}\label{I2smallx}
I^{(2)}(x) \sim \left.\begin{cases} 
2ce^{-2\pi l} \sim 2^{-7}cx^2\,,\qquad & \hat d>1 \\
(2c+\hat m)e^{-2\pi l} \sim 2^{-8}(2c+\hat m)x^2\,,\qquad & \hat d=1 \\
\hat me^{-2\pi\hat d l} \sim \hat m(x/16)^{2\hat d}\,,\qquad & \hat d<1
\end{cases}\right\} \qquad
(x\ll1)\,.
\end{equation}
This term can be matched onto the leading term in the expansion in intermediate states \eqref{MRIOPE}, by noting that the lowest-dimension operator of $\CC^2/\Z_2$ appearing in the $\sigma\sigma$ OPE, other than the unit operator, is $\hat{\mathcal{A}}\otimes\hat{\mathcal{A}}$ if $\hat d\le1$, and the stress tensor if $\hat d\ge1$. The OPE coefficients are computed and matched to \eqref{I2smallx} in Appendix \ref{C2/Z2} (see also the discussion around \eqref{stress}), where we also consider more generally the matching between the Lunin-Mathur formula for the four-point function \eqref{4point} and its expansion in intermediate states.

\FIGURE{
\includegraphics[width=2.5in]{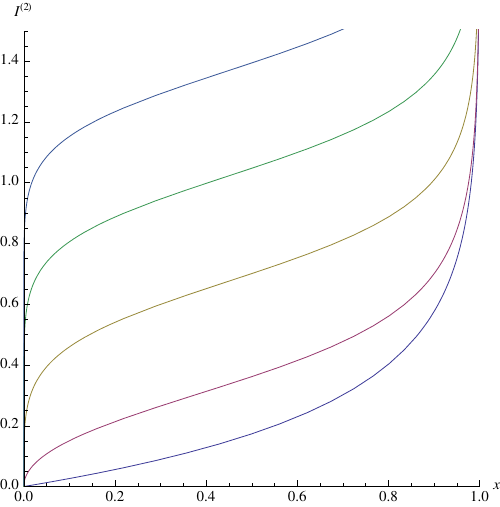}
\caption{Mutual R\'enyi information of intervals $[0,x]$ and $[1,\infty]$ for a free scalar field of radius $R$, with (from bottom to top) $R^2 = 1,2,4,8,16$.
}
\label{I2vsxfree}
}

A simple example of the application of \eqref{MCI} is to a free scalar compactified on a circle of radius $R$. The torus partition function is
\begin{equation}
Z_{il} = \frac{\theta_3(il/R^2)\theta_3(ilR^2)}{\eta^2(il)}\,,
\end{equation}
so \cite{Furukawa:2008uk,Calabrese:2009ez}
\begin{equation}
I^{(2)}(x) =\ln\left(\frac{\theta_3(il/R^2)\theta_3(ilR^2)}{\theta_3(il)\theta_4(il)}\right).
\end{equation}
This is plotted against $x$ for several values of $R$ in figure \ref{I2vsxfree}. The small-$x$ behavior is as predicted by \eqref{I2smallx}, with the lowest non-unit operator having dimension $\hat d=1/(2R^2)$ (for $R^2\ge1$) and multiplicity $\hat m = 2$ (except for $R=1$, where the lightest winding and momentum modes are degenerate, so $\hat m=4$) \cite{Furukawa:2008uk,Calabrese:2009ez}.

\subsubsection{Holographic CFTs}

We now turn to holographic CFTs, briefly reviewing Maldacena and Strominger's result for the torus partition function \cite{Maldacena:1998bw}. Expanding the free energy in powers of $c^{-1}\sim G_{\rm N}$, the leading term is of order $c$ and is given by the Euclidean action of the dominant saddle point. Here the boundary condition is simply that the conformal boundary should be the torus with $\tau = il$; there are no operators inserted in the path integral so no fields other than the metric are sourced. For $l>1$ ($x<1/2$) the dominant saddle point is the Euclidean BTZ black hole, which is topologically a solid torus in which the Euclidean time circle (the circle of length $l$) is contractible. For $l<1$ ($x>1/2$) the dominant saddle point is Euclidean AdS${}_3$ with the Euclidean time direction periodically identified; the topology is a solid torus in which the spatial circle is contractible.\footnote{Note that the contractibility of the two cycles of the boundary torus requires anti-periodicity of fermions on both. As explained in footnote \ref{torusfermions}, these are precisely the boundary conditions we have in this case.} The phase transition between the two saddles, the Hawking-Page transition, is first-order, so the free energy, and hence $I_1^{(2)}(x)$, has a discontinuous first derivative. Specifically, the Euclidean actions of the two saddle points yield \cite{Maldacena:1998bw}
\begin{equation}\label{holoZ}
\ln Z_{il} =\begin{cases} 2\pi c/(12l)+O(c^0)\,,&\qquad l<1\\  2\pi cl/12+O(c^0)\,,&\qquad l>1\end{cases}\,,
\end{equation}
so
\begin{equation}\label{holoMRI}
I^{(2)}_1(x) = 
\ln\left(\frac{\theta_2(il)}{2\eta(il)}\right)+ 
\begin{cases} 2\pi/(12l)\,,&\qquad l<1\\  2\pi l/12\,,&\qquad l>1\end{cases}\,,
\end{equation}
which is plotted in figure \ref{I2vsxholo}. Note that, although $I^{(2)}_1(x)$ does not vanish in the region $x<1/2$, it is numerically quite small---smaller than $I^{(2)}(x)$ for the free scalar by two orders of magnitude or more. The expansion for small $x$ is $2^{-7}x^2$; comparing to \eqref{I2smallx}, it is as if the lowest-dimension operator of $\CC^2/\Z_2$ is the stress tensor. In fact, there are other operators, but since their multiplicity is finite in the limit $c\to\infty$, they do not contribute to $I^{(2)}_1$. We will discuss this expansion in detail in subsection \ref{expansionx}.

\FIGURE{
\includegraphics[width=2.8in]{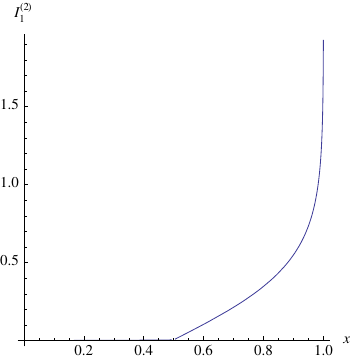}\qquad
\includegraphics[width=2.8in]{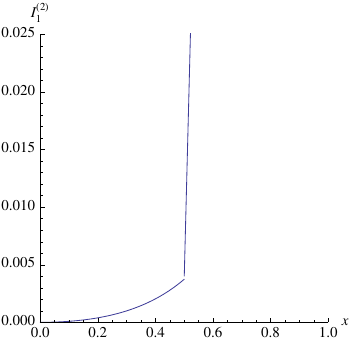}
\caption{Coefficient of $c$ in the mutual R\'enyi information of intervals $[0,x]$ and $[1,\infty]$ in a general holographic CFT. The two plots differ only in the scale of the vertical axis. In particular, the plot on the right shows that $I_1^{(2)}$ is non-zero (although quite small) for $x<1/2$.
}
\label{I2vsxholo}
}

If we compare the bulk saddle-point geometries used to derive \eqref{holoZ} to the ones obtained from Fursaev's construction, we see that they are topologically identical but metrically different. To describe the geometry obtained from Fursaev's construction, we add a Euclidean time direction, coming out of the page, to each diagram in figure \ref{minimal}, and consider the double cover of the resulting three-dimensional geometry, branched over $r_{\rm dis}$ and $r_{\rm con}$ respectively. On the left-hand diagram, relevant when $x<1/2$, the cycle on the boundary that encircles $u_1$ and $v_1$, staying on one sheet, is contractible through the bulk; this is the spatial circle of the torus. On the right-hand diagram, relevant when $x>1/2$, the boundary cycle that encircles $v_1$ and $u_2$, crossing both branch cuts, is contractible through the bulk; this is the time circle of the torus. Hence in each case the topology is precisely the same as that of the true saddle-point geometry. However, their metrics are different; in particular, while the former are singular, the latter are smooth.

\subsubsection{Large-$c$ CFTs}

The formula \eqref{holoZ} for the torus partition function, and therefore the formula \eqref{holoMRI} for the MRI, applies not only to holographic CFTs but also to symmetric-product theories $\CC=\CC_0^N/S_N$ at large $N$, where $\CC_0$ is any (compact unitary) CFT \cite{Dijkgraaf:2000fq,MinwallaRaju}.\footnote{We thank S. Minwalla for helpful discussions on the material in this subsection.} The basic reason is that, for $l<1$, the effective temperature for the long strings, which dominate the partition function, is enhanced by a factor of $N$, so the theory is effectively always in the high-temperature limit; the partition function for $l>1$ is then given by modular invariance. An example is the supersymmetric $(T^4)^N/S_N$ theory, which is conjectured to be connected in a moduli space to type IIB string theory on AdS${}_3\times S^3\times T^4$. It would seem reasonable then to guess that the torus partition function is given by \eqref{holoZ} for all theories on this moduli space. In other words, the MRI appears to enjoy a non-renormalization theorem.

The phase transition at $x=1/2$ can be understood in terms of the expansion $Z_{il} = \sum_ie^{-2\pi(d_i+c/12)l}$, where the $d_i$ are the scaling dimensions of the operators of $\CC$. In any fixed theory, with finite $c$, this expansion converges and is analytic for all $x\in[0,1)$. In the large-$c$ limit, the operators of $\CC$ can (roughly speaking) be divided into those with scaling dimensions of order 1 (``short strings") and those with scaling dimensions of order $c$ (``long strings"). Each long-string operator makes a contribution to the sum that is exponentially suppressed in $c$. However, the number of long-string operators is exponentially large in $c$, so they may actually dominate the sum. In fact, whether the short strings or the long strings dominate depends on the value of $l$, and hence of $x$. In both holographic and large-$N$ symmetric-product theories, short strings dominate for $x<1/2$ and long strings for $x>1/2$. Thus the order-$c$ part of the free energy is due entirely to short strings for $x<1/2$ and to long strings for $x>1/2$.

\subsection{MRI for $n>2$}\label{MRIng2}

In this subsection we will extend our study of $I^{(n)}(x)$ to larger values of $n$. Although we will not be able to give explicit formulas, we will argue that all the main qualitative features carry through. In particular, the existence for all $n>1$ of a discontinuous first derivative in $I^{(n)}_1(x)$ at $x=1/2$ constitutes significant evidence in favor of the RT formula, which predicts precisely such a phase transition for $n=1$.

According to equation \eqref{MRI2intervals}, the MRI $I^{(n)}(x)$ for general $n$ is given in terms of the four-point function of twist operators in the orbifold theory $\CC^n/\Z_n$. This four-point function is in turn equal to the (renormalized) zero-point function of $\CC$ on the surface $E_n$, made of $n$ sheets connected by a branch cut running from $0$ to $x$ and another from $1$ to $\infty$. This is the Riemann surface for the equation $y^n = z(z-1)/(z-x)$, which has genus $n-1$ and a complex structure that depends on $x$. As in the genus-1 case studied in the previous subsection, the surface $E_n$ can be taken by a Weyl transformation to a fiduciary metric with the same complex structure, for example the constant-curvature metric. Hence there is an analogue of \eqref{MRI2intervals}, in which $I^{(n)}$ is written as a sum two terms:
\begin{equation}
I^{(n)}(x) = \ln Z^{(n)}_x + cI_{1,\rm geometric}^{(n)}(x)\,.
\end{equation}
$Z^{(n)}_x$ is the partition function of $\CC$ on the surface carrying the fiduciary metric. The second term is derived from the Liouville action for the Weyl transformation from $E_n$ to the fiduciary metric; aside from the overall coefficient $c$, it is independent of the particular theory $\CC$, giving a universal contribution to $I^{(n)}_1$.

The geometrical term $I^{(n)}_{1,\rm geometric}$ has not been explicitly computed for $n>2$. However, assuming that the fiduciary metric is chosen to depend smoothly on $x$ (as does, for example, the constant-curvature metric), the Weyl transformation and hence $I^{(n)}_{1,\rm geometric}$ will be smooth functions of $x$.

Meanwhile, the genus-$(n-1)$ partition function $Z_x^{(n)}$ is known explicitly for $n>2$ only in a small number of CFTs. For holographic theories, despite considerable progress (especially in the context of pure gravity theories), explicit formulas are not available, even in the large-$c$ limit; see for example \cite{Krasnov:2000zq,Witten:2007kt,Yin:2007gv,Yin:2007at,Maloney:2007ud}. Even in the absence of an explicit formula, however, we can argue that the partition function is smooth except at $x=1/2$, where it has a discontinuous first derivative, just as we saw for $n=2$ in the last subsection. It is known that phase transitions, analogous to the Hawking-Page transition, occur at fixed points of the mapping-class group (the group of large diffeomorphisms of the Riemann surface). The reason is that different saddle points of the bulk gravitational action are effectively mapped onto each other by the action of the mapping-class group, and therefore at a fixed point they necessarily have the same action.\footnote{We thank A.\ Maloney for helpful discussions on this point.} On the surface $E_n$, there is an element of the mapping-class group that effectively takes $x$ to $1-x$; for example, the surfaces $E_n$ with $x=1/3$ and $x=2/3$ have the same complex structure up to the action of an element of the mapping-class group. That element permutes the cycle that encircles $0$ and $x$ with the one that encircles $x$ and $1$. It has a unique fixed point, namely $x=1/2$. Hence we expect a phase transition in (the order-$c$ part of) $\ln Z_x^{(n)}$, and therefore in $I_1^{(n)}(x)$, at $x=1/2$, and only there. This is precisely the property predicted by the RT formula for $I^{(1)}_1(x)$ (see \eqref{I11}). (Furthermore, it seems likely that, as we saw for $n=2$, the dominant saddle has the same topology as predicted by Fursaev, i.e\ for $x<1/2$ the cycle encircling $0$ and $x$ is contractible in the bulk, while for $x>1/2$ the cycle encircling $x$ and $1$ is contractible in the bulk.)

In the previous subsection we used the fact that the torus partition function is the same (at leading order in $c$) for large-$N$ symmetric-product theories ($\CC_0^N/S_N$) as for holographic ones, to support the claim of universality (theory-independence) for $I_1^{(2)}(x)$. It would be interesting to investigate whether the same holds for $Z_x^{(n)}$ for $n>2$. Our results in the next subsection, which apply for all $n$, support such universality.

\subsection{Expansion in $x$}\label{expansionx}

In subsection \ref{phasetransition} we derived an explicit expression for $I_1^{(2)}(x)$ that applied to both holographic and large-$N$ symmetric-product theories. Unfortunately, as we discussed in the last subsection, the computation of higher-genus partition functions, and therefore $I^{(n)}_1(x)$, remains out of reach technically in such theories. Even if we could find explicit expressions, their analytic continuation to general $\alpha$, and in particular to $\alpha=1$, may not be feasible.\footnote{We remind the reader that throughout this paper we use $n$ to denote a positive integer and $\alpha$ a non-negative real number. We assume that all quantities are analytic functions of $\alpha$.} (For example, Calabrese, Cardy, and Tonni were able to compute $I^{(n)}(x)$ for $n>1$ for a compactified free boson \cite{Calabrese:2009ez}. However, the analytic continuation of the resulting expression, which involves a Riemann-Siegel theta function, is unknown.) In this subsection, therefore, we will take a different approach, and compute $I^{(n)}_1(x)$ order by order in $x$ for general $n>1$. The coefficient of each power of $x$ will be a simple enough function of $n$ to allow us to analytically continue it straightforwardly.\footnote{This procedure was applied to the compactified free boson in \cite{Calabrese:2009ez}.} As discussed in the previous subsection, for each $n>1$, $I^{(n)}_1(x)$ is analytic on $[0,1/2]$, but not on larger intervals. Assuming that this property continues to hold for $I^{(\alpha)}_1(x)$ for general $\alpha$, the formulas we derive will be valid on that interval. We are particularly interested in testing two hypotheses concerning the coefficient of each term in the power-series expansion of $I^{(\alpha)}_1(x)$: that it is ``universal" in the sense of subsection \ref{universality}, i.e.\ the same for all theories in the class we are considering; and that it goes to zero in the limit $\alpha\to1$, in agreement with the RT prediction \eqref{I11}. We will find significant evidence in favor of both hypotheses.

\subsubsection{Set-up}\label{set-up}

We are considering theories $\CC$ for which the number of operators with any given dimension is finite in the large-$c$ limit. This condition applies to holographic and $\CC_0^N/S_N$ theories (where $\CC_0$ is held fixed as we take $c\sim N\to\infty$), but not for example to the theory $\CC_0^N$ without the orbifold. We will also assume that the $n$-point functions of primaries in $\CC$ do not diverge in the large-$c$ limit; again, this property holds for holographic and $\CC_0^N/S_N$ theories.

We begin with the relation \eqref{MRIOPE}, in the form
\begin{equation}\label{MRIOPE2}
\exp\left((n-1)I^{(n)}(x)\right)
= \sum_m{c^{\sigma_1}}_{\sigma_1m}{c^m}_{\sigma_1\sigma_{-1}}x^{d_m}\,.
\end{equation}
At this stage we have not taken the large-$c$ limit, and the convergence of the OPE dictates that the sum on the right-hand side converges and is analytic for all $x\in[0,1)$. Only untwisted-sector operators of $\CC^n/\Z_n$ occur in the $\sigma_1\sigma_{-1}$ OPE; these are of the form $\mathcal{A}'_m = (\mathcal{A}_1\otimes\cdots\otimes\mathcal{A}_n)_{\rm sym}$, where the $\mathcal{A}_i$ are operators of $\CC$ (``sym" means average over cyclic permutations).

Before considering the large-$c$ limit, it is interesting to ask what the assumed analyticity in $\alpha$ of $I^{(\alpha)}(x)$ implies about the coefficients on the right-hand side of \eqref{MRIOPE2}. First, except for the leading term, 1, the total coefficient of each power of $x$ must be a multiple of $n-1$ (i.e.\ when analytically continued must have a root at $n=1$). Although we don't know a CFT proof of this statement, it can be tested to any given order. For example, for the holomorphic and antiholomorphic parts of the stress tensor, $T'$ and $\tilde T'$, we have
\begin{equation}
{c^{\sigma_1}}_{\sigma_1T'} = {c^{\sigma_1}}_{\sigma_1\tilde T'} = \frac{d_\sigma}2\,.
\end{equation}
The Zamolodchikov metric on these operators is $\mathcal{G}_{T'T'} = \mathcal{G}_{\tilde T'\tilde T'} = nc/2$, since the central charge of $\CC^n/\Z_n$ is $nc$; they do not mix with each other or with other operators, so $\mathcal{G}^{T'T'} = \mathcal{G}^{\tilde T'\tilde T'} = 2/(nc)$, and
\begin{equation}
{c^{T'}}_{\sigma_1\sigma_{-1}} = {c^{\tilde T'}}_{\sigma_1\sigma_{-1}} = \frac{d_\sigma}{nc}\,.
\end{equation}
Hence their contribution to the sum on the right-hand side of \eqref{MRIOPE2} is $2ax^2$, where
\begin{equation}\label{stress}
a \equiv \frac{d_\sigma^2}{2nc} = \frac{(n^2-1)^2c}{288n^3}\,,
\end{equation}
which has the required factor of $n-1$. The contribution of operators consisting of two identical scalar primaries $\OO$ of $\CC$, with the rest of the $\mathcal{A}_i$ the identity, such as $(\OO\otimes\OO\otimes I\otimes\cdots\otimes I)_{\rm sym}$, is computed in Appendix \ref{OPEprimary}, and it is shown that (at least for half-integer values of the scaling dimension of $\OO$) the analytic continuation again has a zero at $n=1$. More generally, analyticity of $I^{(\alpha)}(x)$ demands that the contribution of an operator made up of $k$ non-unit operators of $\CC$, such as $(\mathcal{A}_1\otimes\cdots\otimes\mathcal{A}_k\otimes I\otimes\cdots\otimes I)_{\rm sym}$, should contain a factor $(n-1)(n-2)\cdots(n-k+1)$, simply because that operator only exists for $n\ge k$.

\subsubsection{Large-$c$ limit}\label{largec}

We now consider the large-$c$ limit. Since we are working order by order in $x$, we will consider only ``short-string" operators of $\CC$, i.e.\ those whose scaling dimensions are finite in the large-$c$ limit. A convenient machinery for systematically computing terms in the sum \eqref{MRIOPE2} is provided by the conformal blocks. We thus write it as a sum over primaries $\mathcal{O}'_m$ (of $\CC^n/\Z_n$):
\begin{equation}\label{MRIOPE3}
\exp\left((n-1)I^{(n)}(x)\right)
= \sum_mC_m\mathcal{F}(h_m,x)\mathcal{F}(\tilde h_m,x)x^{h_m+\tilde h_m}\,,
\end{equation}
where $C_m \equiv {c^{\sigma_1}}_{\sigma_1m}{c^m}_{\sigma_1\sigma_{-1}}$, $h_m$ and $\tilde h_m$ are the weights of $\mathcal{O}'_m$, and $\mathcal{F}(h_m,x)$ is (up to a factor of  $x^{2d_\sigma-h_m}$) the conformal block with all 4 external operators of weight $h_\sigma = \tilde h_\sigma = d_\sigma/2$. ($C_m$ and the conformal blocks also depend implicitly on $c$ and $n$.) The conformal blocks can be computed straightforwardly, albeit somewhat tediously, to any desired order in $x$. Using \emph{Mathematica}\footnote{For these computations we used the package Virasoro.nb, available at \href{http://people.brandeis.edu/~headrick/physics/}{http://people.brandeis.edu/-~headrick/physics/}.}, we computed them to order $x^5$. In the following expression, the first two terms are exact, while the subsequent ones have been expanded in powers of $a^{-1}\sim c^{-1}$:
\begin{align}\label{block}
\mathcal{F}(h_m,x) &= 1\nonumber \\
&\quad+\frac{h_m}2x \nonumber \\
& \quad + \left(a+f_{20} + O(a^{-1})\right)x^2 \nonumber \\
& \quad + \left(\left(1+\frac{h_m}2\right)a+f_{30} + O(a^{-1})\right)x^3 \nonumber \\
& \quad + \left(\frac12a^2+f_{41}a + f_{40} + O(a^{-1})\right)x^4 \nonumber \\
& \quad + \left(\left(1+\frac{h_m}4\right)a^2+f_{51}a + f_{50} + O(a^{-1})\right)x^5 \nonumber \\
& \quad + O(x^6)\,.
\end{align}
The $f_{ij}$ are rational functions of $h_m$ and $n$ (regular at $n=1$); their precise form is not important for us, except for one feature we will point out below. We now factor out the positive powers of $a$ (i.e.\ of $c$), since those determine $I^{(n)}_1(x)$. It turns out that they organize themselves naturally into an exponential:
\begin{equation}
\mathcal{F}(h_m,x) = F(h_m,x)\exp\left(ax^2+ax^3+g_4 ax^4+ g_5 ax^5+O(x^6)\right)\,,
\end{equation}
where, by definition, $F$ contains only non-positive powers of $a$:
\begin{equation}
F(h_m,x) = 1 + \frac{h_m}2x + \left(f_{20}+O(a^{-1})\right)x^2 + \left(f_{30}+O(a^{-1})\right)x^3 + O(x^4)\,.
\end{equation}
Remarkably, thanks to some cancellations among the $f_{ij}$, the coefficients of $ax^4$ and $ax^5$ turn out to be independent of $h_m$:
\begin{align}\label{betadef}
g_4 &= f_{41}-f_{20}-\frac{h_m}2 = \frac{1309n^4-2n^2-11}{1440n^4} \nonumber \\
g_5 &= f_{51}-f_{30}-f_{20}-g_4\frac{h_m}2 = \frac{589n^4-2n^2-11}{720n^4}\,.
\end{align}
Assuming that this pattern continues to higher orders, it allows us to pull the exponential out of the sum \eqref{MRIOPE3}, and write:
\begin{equation}\label{CBresult}
I^{(n)}_1(x) = J^{(n)}(x) + \frac{(n-1)(n+1)^2}{144n^3}\left(x^2+x^3+g_4 x^4+g_5x^5+O(x^6)\right)\,,
\end{equation}
where $J^{(n)}(x)$ is the contribution to $I^{(n)}_1(x)$ (if any) from the OPE coefficients $C_m$:
\begin{equation}
J^{(n)}(x) \equiv \frac1{n-1}\lim_{c\to\infty}\frac1c\ln\left(\sum_mC_mF(h_m,x)F(\tilde h_m,x)x^{h_m+\tilde h_m}\right).
\end{equation}

Before discussing $J^{(n)}(x)$, let us point out several noteworthy features of the second term of \eqref{CBresult}. First, it does not depend at all on the particular theory, supporting the universality proposed in subsection \ref{universality}; this is a consequence of the cancellation of the $h_m$-dependence in $g_4$ and $g_5$, \eqref{betadef}. Second, if we set $n=2$, it agrees with the expansion to fifth order of \eqref{holoMRI}; hence $J^{(2)}(x)$ vanishes at least to fifth order. Third, it can be straightforwardly continued to non-integer values of $\alpha$, and vanishes at $\alpha=1$:
\begin{equation}\label{CBresult2}
I^{(\alpha)}_1(x) = J^{(\alpha)}(x) + \frac{(\alpha-1)(\alpha+1)^2}{144\alpha^3}\left(x^2+x^3+g_4 x^4+g_5x^5+O(x^6)\right)\,.
\end{equation}
The fact that the second term vanishes at $\alpha=1$, which provides strong quantitative evidence in favor of the RT formula, can be traced to the fact that the conformal block \eqref{block} depends on $c$ through $a\sim (n-1)^2c$.

It remains to ask what we can say about $J^{(n)}(x)$. Since we are disallowing theories in which the number of primaries of a given dimension is proportional to $c$, $J^{(n)}(x)$ will be non-zero if and only if some of the coefficients $C_m$ contain positive powers of $c$. Some of the primaries of $\CC^n/\Z_n$ are products of primaries of $\CC$: $\mathcal{O}'_m = (\mathcal{O}_1\otimes\cdots\otimes\mathcal{O}_n)_{\rm sym}$. For these, as we show in appendix \ref{OPEprimary}, the OPE coefficient ${c^{\sigma_1}}_{\sigma_1m}$ is given by an $n$-point function of the constituent operators $\mathcal{O}_1,\ldots,\mathcal{O}_n$ (see \eqref{genOPE}; this only applies to scalars, but the only change for operators with spin will be the presence of certain phase factors). In holographic and symmetric-product ($\CC_0^N/S_N$) theories, these $n$-point functions go like $c^{1-k/2}$, where $k$ is the number of non-identity operators, so $C_m\sim c^{2-k}$ (this is for $k>1$; for $k=0$, i.e.\ the identity of $\CC^n/\Z_n$, $C_1=1$, while for $k=1$, $C_m=0$, since the one-point function of a non-identity operator vanishes). For example, for $k=2$ we have a two-point function, which is clearly independent of $c$ ($C_m$ is computed in this case in appendix \ref{OPEprimary}). Hence primaries of $\CC^n/\Z_n$ that are products of primaries of $\CC$ do not contribute to $J^{(n)}(x)$. However, there are other primaries of $\CC^n/\Z_n$ that are made up of descendants of $\CC$.\footnote{\label{C2Z2primary}For an example of a primary of $\CC^n/\Z_n$ that is made up of descendants of $\CC$, let $n=2$ and let the Virasoro generators of $\CC^2/\Z_2$ be $L'_m=L_m\otimes I+I\otimes L_m$. Given a non-unit primary $\OO$ of $\CC$, one can construct three linearly independent operators at level 2 out of $\OO$ and its descendants, namely
\begin{align}
\mathcal{A}'_1&=L_{-2}\cdot\OO\otimes\OO+\OO\otimes L_{-2}\cdot\OO \\
\mathcal{A}'_2&=L_{-1}\cdot\OO\otimes L_{-1}\cdot\OO\\
\mathcal{A}'_3&=L_{-1}^2\cdot\OO\otimes\OO+\OO\otimes L_{-1}^2\cdot\OO\,.
\end{align}
However, there are only two descendants of this form, namely $L'_{-2}\cdot(\OO\otimes\OO)=\mathcal{A}'_1$ and $L_{-1}^{\prime2}\cdot(\OO\otimes\OO) = 2\mathcal{A}'_2+\mathcal{A}'_3$. Therefore one linear combination of $\mathcal{A}'_{1,2,3}$ must be primary. (This primary does not contribute to $J^{(2)}(x)$, as we know since the latter vanishes at least to order $x^5$. However, its generalizations for $n>2$ might contribute to $J^{(n)}(x)$.)} The computation of $C_m$ for such operators is more involved, because of the more complicated transformation law for the $\CC$-descendants in going from the $n$-sheeted plane to the standard plane, which can bring in additional powers of $c$. We have not attempted this calculation, and it is possible that such operators contribute to $J^{(n)}(x)$. If so, the conjecture is then that $J^{(\alpha)}(x)$ is independent of the particular theory $\CC$, and vanishes at $\alpha=1$. It should be straightforward in principle to compute $C_m$ and test these conjectures in specific examples.

In this subsection, we have provided non-trivial evidence, based on the expansion \eqref{MRIOPE2}, that $I_1^{(\alpha)}(x)$ is theory-independent and, for $0\le x\le1/2$, vanishes at $\alpha=1$. In view of the pattern we have found, it would clearly be desirable to have some general understanding of the structure of the OPE coefficients at large $c$ that leads to these properties. We leave the exploration of this structure to future work.

\section{Generalizations, open questions, and discussion}\label{discussion}

In the previous section, through the study of R\'enyi entropies, we provided strong evidence in favor of the Ryu-Takayanagi formula. We focused on one of the simplest non-trivial field-theory examples, namely two disjoint intervals in the vacuum of a two-dimensional CFT. Along the way, we found evidence that a large class of large-$c$ theories share the same entanglement (R\'enyi) entropies. It would be interesting to extend our analysis to more general situations, including: more than two intervals; states other than the vacuum, such as thermal states; CFTs on the circle rather than the line; CFTs in more than two dimensions; and non-conformal field theories. In particular, it is clear that the two key qualitative predictions of the RT formula persist in all these examples, namely that there is a phase transition in the mutual information between two regions as a function of their sizes and separations, and that it vanishes on one side of the phase transition. One should be able to test these predictions using similar techniques to the ones used in this paper, namely classical gravity and the OPE. One should also be able to test whether the EREs are the same for non-holographic theories with large central charges.

Our analysis leaves a number of open questions. We will start with the more technical ones, and move towards the more conceptual.

First, our calculation of $I^{(n)}_1(x)$ in subsection \ref{largec} left out the term $J^{(n)}(x)$, which comes from primary operators of $\CC^n/\Z_n$ that are composed of descendants of $\CC$. It would be useful to evaluate this term, at least up to some power of $x$, to confirm both its theory-independence and that it vanishes at $n=1$. More generally, it should be possible to understand on general CFT grounds the pattern found in subsection \ref{largec} that, in the four-point function of twist operators, every factor of $c$ is accompanied by a factor of $(n-1)^2$.

Second, it would be very interesting to compute the MRI $I^{(n)}_1(x)$ explicitly for $n>2$ in holographic and large-$N$ symmetric-product theories, to see, first, if they agree, and second, if they indeed have a phase transition at $x=1/2$. Better yet would be to analytically continue the resulting expressions to general $\alpha$, and directly confirm or refute the RT formula in this case.

Third, we saw that, for $x\le1/2$, the MRI is of order $c$ for $\alpha\neq1$ while the MI is only of order 1. In order to understand the behavior better, it would be interesting to find a simple toy-model system with a large number of degrees of freedom, in which the MRI between two subsystems is of order of the number of degrees of freedom while the MI is only of order $1$.

Lastly, and perhaps most importantly, we should ask what the status of the RT formula is, given the results of this paper. On the one hand, we have provided strong evidence that it is correct. On the other hand, have we understood any better \emph{why} it should be true? In particular, why does the minimal surface play a critical role in the entanglement entropy, and what is the physical significance of the bulk region $r_A$ that it bounds? Fursaev's proof \cite{Fursaev:2006ih}, though incorrect, had the advantage of explaining in a simple and elegant manner the role of the minimal surface. On the other hand, in the R\'enyi entropy calculations we have performed in this paper, this role is not so clear. Rather, the agreement between the R\'enyi entropies and the RT formula appeared to be almost fortuitous. Clearly, while the RT formula provides a tantalizing hint about the structure of quantum information in holographic theories, most of that structure still remains hidden from view.

\acknowledgments

First and foremost, I would like to thank S. Minwalla for initial collaboration on this project; essentially all of the new ideas in this paper originated during discussions with him. It is also a pleasure to thank the following people for helpful discussions: M. Gaberdiel, S. Hartnoll, V. Hubeny, A. Lawrence, L. Levitov, H. Liu, A. Maloney, J. McGreevy, M. Rangamani, T. Takayanagi, E. Tonni, M. van Raamsdonk, and T. Wiseman; and to thank T. Takayanagi, E. Tonni, and M. van Raamsdonk for useful comments on a draft. Finally, I would like to thank the Tata Institute of Fundamental Research, where this project was initiated, and Korea Institute for Advanced Study, where it was completed, for their hospitality. This research was supported in part by DOE grant No.\ DE-FG02-92ER40706.

\appendix

\section{Computations in the $\CC^n/\Z_n$ orbifold theory}

\subsection{Analysis of four-point function of twist operators in $\CC^2/\Z_2$}\label{C2/Z2}

In this appendix, we consider a general modular-invariant, compact, unitary CFT $\CC$ with central charge $c$, and its symmetric square $\CC^2/\Z_2$. (At the end we also make some comments about the orbifold theory $\CC^n/\Z_n$ for general $n$.) The orbifold theory has central charge $2c$, and contains a single twist operator $\sigma$ with conformal weights $h_\sigma = \tilde h_\sigma = c/16$. Lunin and Mathur \cite{Lunin:2000yv} computed the four-point function of these operators, showing that it is determined by the partition function $Z_\tau$ of $\CC$ on a flat torus with modular parameter $\tau$,
\begin{equation}\label{LM}
\bev{\sigma(0)\sigma(x)\sigma(1)\sigma'(\infty)} = \left|2^8x(1-x)\right|^{-c/12}Z_\tau\,,
\end{equation}
where $\tau$ and $x$ are related by\footnote{In terms of Lunin and Mathur's variables, $x=1/w$ and $\tau = -1/\tau_{\text{Lunin-Mathur}}$. In the bulk of the paper we consider $x$ to be real and lying in the interval $0<x<1$, but in this appendix we will let $x$ be a general complex number.  As in the main text, $\sigma_1'(\infty)\equiv\lim_{z\to\infty}z^{d_\sigma}\sigma_1(z)$.}
\begin{equation}
x = \frac{\theta_2^4(\tau)}{\theta_3^4(\tau)}\,.
\end{equation}
The reason for the appearance of the torus partition function of $\CC$ is that the four-point function of twist operators is the (renormalized) zero-point function on the two-sheeted Riemann surface $E_2$ with a branch cut connecting $0$ to $x$ and another one connecting $1$ to $\infty$. $E_2$ is a torus with complex structure $\tau$. The Weyl transformation that flattens it leads to the prefactor $|2^8x(1-x)|^{-c/12}$.

Both sides of \eqref{LM} can be decomposed into a sum of states, and we would like to understand the relationship between these two decompositions. The torus partition function is a sum over states $\mathcal{A}_m$ in $\CC$:\footnote{For simplicity we are taking all states to be bosonic.}
\begin{equation}\label{Ztau}
Z_\tau = \sum_mq^{h_m-c/24}\bar q^{\tilde h_m-c/24} = \sum_i\chi_{c,h_i}(q)\chi_{c,\tilde h_i}(\bar q)\,,
\end{equation}
where $q\equiv e^{2\pi i\tau}$. In the second equality, we have grouped the states into conformal families. Each family is labelled by its primary operator $\OO_i$, and $\chi_{c,h_i}$ is its Virasoro character:
\begin{equation}\label{characterdef}
\chi_{c,h_i}(q) = q^{-c/24+h_i}\sum_{N=0}^\infty d(N)q^N\,,
\end{equation}
where $d(N)$ is the number of descendants of $\mathcal{O}_i$ at level $N$. The decomposition \eqref{Ztau} can be obtained by cutting the torus along a cycle and inserting a complete set of states. In the usual presentation of the torus as $\C/(\Z+\tau\Z)$, that cycle should be horizontal.

Meanwhile, the left-hand side of \eqref{LM} can be written as a sum over intermediate states $\mathcal{A}_l'$ of $\CC^2/\Z_2$, with weights $(h'_l,\tilde h'_l)$:
\begin{equation}\label{4pointdecomp}
\bev{\sigma(0)\sigma(x)\sigma(1)\sigma'(\infty)} =
\sum_lc_{\sigma\sigma l}{c^l}_{\sigma\sigma}x^{h'_l-c/8}\bar x^{\tilde h'_l-c/8}\,,
\end{equation}
where $c_{\sigma\sigma l} = \ev{\sigma'(\infty)\sigma(1)\mathcal{A}'_l(0)}$ and ${c^l}_{\sigma\sigma}$ is the coefficient of $\mathcal{A}'_l$ in the $\sigma$--$\sigma$ OPE. Assuming for clarity that $|x|<1$, this decomposition is obtained by cutting the sphere on a circle of radius $r$ ($|x|<r<1$) around the origin, which separates the twist operators located at 0 and $x$ from those located at $1$ and $\infty$, and inserting a complete set of states.

The intermediate states in \eqref{4pointdecomp} can also be organized into conformal families, leading to a sum of conformal blocks. However, since the conformal families of $\CC^2/\Z_2$ are not in one-to-one correspondence with the conformal families of $\CC$ (see footnote \ref{C2Z2primary}), and we are trying to reproduce the sum \eqref{Ztau} which is over the latter, we will organize the intermediate states slightly differently. First we note that only untwisted states appear in the sum, and these are of the form $\mathcal{A}'_l = \mathcal{A}_m\otimes\mathcal{A}_n + \mathcal{A}_n\otimes\mathcal{A}_m$, where $ \mathcal{A}_m,\mathcal{A}_n$ are states of $\CC$. We are inserting this state on the circle of radius $r$ mentioned in the previous paragraph. In the presence of the twist operators, we can consider that we are working in the theory $\CC$ on the Riemann surface $E_2$, where the circle is two circles, one on each sheet; we are inserting $\mathcal{A}_n$ on one circle and  $\mathcal{A}_m$ on the other. These two circles both represent the same cycle of the torus, namely the horizontal cycle mentioned below \eqref{characterdef}. In other words we have cut the torus into two finite cylinders. Each cylinder has $\mathcal{A}_n$ inserted on one boundary and $\mathcal{A}_m$ inserted on the other.

We now perform the Weyl transformation that turns $E_2$ into the flat torus. Two things will happen. First, we get the geometrical factor $|2^8x(1-x)|^{-c/12}$, as computed by Lunin and Mathur, which is independent of the states. Second, each state gets mapped by the action of the conformal group to a linear combination of states. By definition, this group acts within conformal families. Hence if $\mathcal{A}_n$ and $\mathcal{A}_m$ are not in the same family, then the cylinder amplitude vanishes. So we can gather the terms in \eqref{4pointdecomp} into conformal families of $\CC$:
\begin{equation}\label{Ksum}
\bev{\sigma(0)\sigma(x)\sigma(1)\sigma'(\infty)} =
\sum_iK_i(x,\bar x)\,,
\end{equation}
where
\begin{equation}\label{Kidef}
K_i(x,\bar x) = \sum_{\text{$\mathcal{A}_m,\mathcal{A}_n$ descendants of $\OO_i$}}
c_{\sigma\sigma(m,n)}{c^{(m,n)}}_{\sigma\sigma}x^{h_m+h_n-c/8}\bar x^{\tilde h_m+\tilde h_n-c/8}\,.
\end{equation}
(The set of operators in $\CC^2/\Z_2$ of the form $\mathcal{A}_m\otimes\mathcal{A}_n + \mathcal{A}_n\otimes\mathcal{A}_m$ where $\mathcal{A}_m$ and $\mathcal{A}_n$ are both descendants of the primary $\OO_i$ in $\CC$, is the union of several conformal families of $\CC^2/\Z_2$. Hence $K_i$ includes several conformal blocks of $\CC^2/\Z_2$.) Each term of \eqref{Ksum} corresponds to precisely one term in the sum on the right-hand side of \eqref{Ztau}, and the Lunin-Mathur formula tells us that
\begin{equation}\label{Ki}
K_i(x,\bar x) = \left|2^8x(1-x)\right|^{-c/12}\chi_{c,h_i}(q)\chi_{c,\tilde h_i}(\bar q)\,.
\end{equation}
It is interesting that $K_i$ factorizes as a holomorphic times an antiholomorphic function.

Each state in the sum \eqref{Kidef} contributes to $K_i$ a monomial in $x,\bar x$, while each state in the sum \eqref{characterdef} contributes to $\chi_{c,h_i}(q)$ a monomial in $q$. The complicated mixing between states due to the action of the conformal group is reflected in the complicated relationship between $x$ and $q$. However, the leading terms for small $x$ on the two sides of \eqref{Ki} can be matched easily. On the right-hand side the leading term is due to the primary $\OO_i$ itself, so we have
\begin{equation}\label{smallx}
\left|2^8x\right|^{-c/12}q^{-c/24+h_i}\bar q^{-c/24+\tilde h_i} \approx 2^{-8h_i-8\tilde h_i}x^{-c/8+2h_i}\bar x^{-c/8+2\tilde h_i}\,,
\end{equation}
where we used the expansion for small $x$, $q \approx 2^{-8}x^2$. The leading term on the left-hand side is due to the operator $\OO_i' = \OO_i\otimes\OO_i$, which has weights $(h_i',\tilde h_i') = (2h_i,2\tilde h_i)$. $\OO'_i$ is primary, so (taking it to be normalized in the Zamolodchikov metric) we have $c_{\sigma\sigma i'} = {c^{i'}}_{\sigma\sigma} = \ev{\sigma(0)\OO'_i(1)\sigma'(\infty)}$. To evaluate this three-point function, we consider the theory $\CC$ on the two-sheeted Riemann surface with a branch cut extending from 0 to $\infty$, and with $\OO_i$ inserted at the point $z=1$ on both sheets. We can use the map $z=t^2$ to relate this to the two-point function $\ev{\OO_i(-1)\OO_i(1)}$ in the $t$-frame. (The factor arising from the Weyl transformation is absorbed in the renormalization of the twist fields.) All in all we find
\begin{equation}
c_{\sigma\sigma i'} = {c^{i'}}_{\sigma\sigma} = 2^{-4h_i-4\tilde h_i}\,,
\end{equation}
which leads immediately to agreement with \eqref{smallx}.

In principle equation \eqref{Ki} can be checked to higher orders. Consider, for example, the conformal family of the identity. For convenience, let us divide both sides of \eqref{Ki} by the leading term:
\begin{equation}\label{K1}
|x|^{c/4}K_1(x,\bar x) = 
\left|
\left(2^8\frac{1-x}{x^2}\right)^{-c/24}
\chi_{c,0}(q)
\right|^2\,.
\end{equation}
Generically, the conformal family of the identity is a full Verma module except the states $L_{-1}|0\rangle$ and $\tilde L_{-1}|0\rangle$ and their would-be descendants, which vanish. (For the minimal models there are also other missing states.) In that case the character is
\begin{equation}
\chi_{c,0}(q) = q^{-c/24}\prod_{n=2}^\infty\frac1{1-q^n} = q^{-c/24}\frac{q^{1/24}(1-q)}{\eta(q)}\,,
\end{equation}
so the holomorphic part of \eqref{K1} is
\begin{equation}
\left(2^8\frac{1-x}{x^2}q\right)^{-c/24}\prod_{n=2}^\infty\frac1{1-q^n}\,.
\end{equation}
The first few terms in the expansion in powers of $x$ are:
\begin{equation}
1+2^{-8}cx^2+2^{-8}cx^3+2^{-17}(c^2+465c+2)x^4\,.
\end{equation}
In the expansion in states of $\CC^2/\Z_2$, \eqref{Kidef}, the quadratic term is due to the stress tensor, while the cubic term is due to $L_{-3}|0\rangle\cong\partial T$. The correct matching of the coefficient for the former can be seen by setting $n=2$ in \eqref{stress}.

Modular invariance means that the torus partition function can be written as a sum of characters in a different way, namely
\begin{equation}\label{Ztau2}
Z_\tau = \sum_i\chi_{c,h_i}(\hat q)\chi_{c,\tilde h_i}(\bar {\hat q})\,,
\end{equation}
where $\hat q\equiv e^{-2\pi i/\tau}$. This decomposition is produced by cutting the torus along its ``vertical" cycle. Meanwhile, associativity of the OPE means that the four-point function of twist operators can be decomposed in intermediate states with each state contributing a power of $1-x$ (instead of $x$ as in \eqref{4pointdecomp}), by cutting along a circle centered on $1$ that separates $1$ and $x$ from $0$ and $\infty$. That circle corresponds to two circles on $E_2$, both representing the vertical cycle. Thus the two decompositions can be mapped to each other just as we did above. It is interesting that the associativity of the OPE in $\CC^2/\Z_2$ is directly related to the modular invariance of $\CC$.

If we attempt to generalize the above analysis to the analogous four-point function of twist operators
\begin{equation}\label{gen4point}
\bev{\sigma_1(0)\sigma_{-1}(x)\sigma_1(1)\sigma'_{-1}(\infty)} 
\end{equation}
in the $\CC^n/\Z_n$ orbifold theory, the following structure emerges. The Riemann surface $E_n$ has $n$ sheets joined by a branch cut extending from $0$ to $x$ and another one extending from $1$ to $\infty$. This surface has genus $n-1$, and the circle centered on 0, that separates the twist operators located at $0$ and $x$ from those located at $1$ and $\infty$, decomposes into $n$ circles, which separate $E_n$ into two $n$-punctured spheres. Again, only untwisted states, which are of the form $(\mathcal{A}_{m_1}\otimes\cdots\otimes\mathcal{A}_{m_n})_\text{sym}$, enter in the sum we insert on that circle. We are left with a sum of squares of $n$-point functions of $\CC$ (to be contrasted with \eqref{4pointdecomp}, which is a sum of squares of three-point functions of $\CC^2/\Z_2$, or in this case $\CC^n/\Z_n$). Unlike in the $\CC^2/\Z_2$ case, the $\mathcal{A}_{m_i}$ do not all have to belong to the same conformal family of $\CC$ to contribute to this sum. For this reason, this decomposition is less immediately useful than for the case $n=2$. There will also be an overall geometrical factor coming from the appropriate Weyl transformation.

\subsection{Computation of certain OPE coefficients}\label{OPEprimary}

In this appendix we will consider primary operators in the orbifold theory  $\CC^n/\Z_n$, of the form
\begin{equation}\label{genOprime}
\OO'_m = \left(\OO_1\otimes\cdots\otimes\OO_n\right)_{\rm sym}\,,
\end{equation}
where the $\OO_i$ are scalar primaries of $\CC$, and the subscript ``sym" implies an average over cyclic permutations. We will first show that ${c^{\sigma_1}}_{\sigma_1m}$, its OPE coefficient with the twist operator $\sigma_1$ onto $\sigma_1$, is given in terms of the $n$-point function in $\CC$ of the component operators $\OO_i$. We will then focus on the simplest non-trivial case, with only two non-identity operators (necessarily the same, for ${c^{\sigma_1}}_{\sigma_1m}$ to be non-zero), such as $(\OO\otimes\OO\otimes I\otimes\cdots\otimes I)_{\rm sym}$, $(\OO\otimes I\otimes\OO\otimes I\otimes\cdots\otimes I)_{\rm sym}$, etc. For a given $\OO$, all such operators have the same dimension $d_m=2d$, where $d$ is the dimension of $\OO$, so in the sum \eqref{MRIOPE2} they all contribute to the coefficient of $x^{2d}$. We compute their total contribution, and show that, when $d$ is an integer, its analytic continuation in $n$ vanishes at $n=1$; this is the property discussed after \eqref{stress}.

We begin with the more general operator \eqref{genOprime}. We compute:
\begin{align}\label{genOPE}
{c^{\sigma_1}}_{\sigma_1m} &= c_{\sigma_{-1}m\sigma_1}\nonumber \\
&= \bev{\sigma_{-1}(0)\OO_m'(1)\sigma'_1(\infty)}_{\CC^n/\Z_n} \nonumber\\
&= \bev{\sigma^\epsilon_{-1}(0)\sigma^\epsilon_1(1)}^{-1}_{\CC^n/\Z_n}
\bev{\sigma^\epsilon_{-1}(0)\OO_m'(1)\sigma^{\epsilon\prime}_1(\infty)}_{\CC^n/\Z_n} \nonumber\\
&= \bev{\sigma^\epsilon_{-1}(0)\sigma^\epsilon_1(1)}^{-1}_{\CC^n/\Z_n}
\left(\bev{\OO_1(e^{2\pi i})\OO_2(e^{4\pi i})\cdots\OO_n(e^{2\pi in})}_{\CC\text{ on }E_n}\right)_{\rm sym} \nonumber\\
&= n^{-\sum_id_i}\bev{\OO_1(e^{2\pi i/n})\OO_2(e^{4\pi i/n})\cdots\OO_n(1)}_{\CC}\,.
\end{align}
(All correlators except the one marked ``$\CC$ on $E_n$" are evaluated on the Riemann sphere.) In the first line we used the fact that the twist operators are normalized, and both they and $\OO'_m$ are primary. In the fourth we used the definition of the twist operators to move to the original theory $\CC$ on the $n$-sheeted surface $E_n$, where the sheets are connected by a branch cut running from 0 to $\infty$ (the positions of the twist operators). The operator $\OO_j$ is positioned at 1 on the $j$th sheet, denoted $e^{2\pi ij}$. In the last line we conformally mapped $E_n$ to the plane by $t=z^{1/n}$. The geometrical factor from the associated Weyl tranformation is independent of the operator insertions, cancelling the factor $\ev{\sigma^\epsilon_{-1}(0)\sigma^\epsilon_1(1)}^{-1}_{\CC^n/\Z_n}$. The operator positions are mapped to the $n$th roots of unity. Since each $\OO_i$ is primary, under the conformal transformation it becomes, in the $t$-frame, $|\partial z/\partial t|^{-d_i}\OO_i = n^{-d_i}\OO_i$. Finally, in the last line the symmetrization was dropped, since a cyclic permutation of the operators is equivalent to a rotation of the plane by $e^{2\pi i/n}$, which leaves the correlator unchanged (all the operators being scalars).

We now specialize to an operator containing exactly two non-identity primaries. Applying \eqref{genOPE} will result in a two-point function of the two operators; in order to get a non-zero result they must therefore be identical:
\begin{equation}\label{specOprime}
\OO'_j = \OO\otimes I^{\otimes (j-1)}\otimes\OO\otimes I^{\otimes(n-j-1)}\,,\qquad1\le j\le\frac n2\,.
\end{equation}
From \eqref{genOPE} we obtain
\begin{equation}
{c^{\sigma_1}}_{\sigma_1j} = n^{-2d}\bev{\OO(e^{2\pi i/n})\OO(e^{2\pi i(j+1)/n})}_\CC
=\left(2n\sin\frac{\pi j}n\right)^{-2d}
\end{equation}
(where $d$ is the dimension of $\OO$). The Zamolodchikov metric for this operator is $\mathcal{G}_{jj}=1/n$, except if $j=n/2$, in which case it is $\mathcal{G}_{jj}=2/n$. Hence we have
\begin{equation}
C_j = {c^{\sigma_1}}_{\sigma_1j}{c^j}_{\sigma_1\sigma_{-1}} = 2^{-\delta_{j,n/2}}n^{1-4d}\left(2\sin\frac{\pi j}n\right)^{-4d}\,.
\end{equation}
The total contribution of these operators to the coefficient of $x^{2d}$ in the sum \eqref{MRIOPE2} is thus
\begin{equation}\label{specOPE}
C_{\rm tot} = \sum_jC_j = \frac{n^{1-4d}}2\sum_{j=1}^{n-1}\left(2\sin\frac{\pi j}n\right)^{-4d}\,.
\end{equation}
We wish to analytically continue this expression in $n$.\footnote{The analytic continuation of the sum in \eqref{specOPE} was also considered in \cite{Calabrese:2009ez}. In particular, an expression was derived that allowed numerical approximations to be computed.} We were not able to do this for general dimension $d$, but in the next paragraph we will show that, when $2d$ is an integer, the sum in \eqref{specOPE} is a polynomial in $n$ of degree $4d$, with a root at $n=1$. This was the statement that was used in subsection \ref{set-up}.

In order to analytically continue the sum in \eqref{specOPE}, we note that the summand equals the reside of the pole at $t=e^{2\pi ij/n}$ of the function
\begin{equation}
f(t) = \frac n{t(1-t)^{2d}(1-t^{-1})^{2d}(t^n-1)}\,.
\end{equation}
We are assuming that $2d$ is a positive integer, so $f(t)$ is single-valued and regular everywhere on the Riemann sphere except for a pole at each $n$th root of unity. In particular, at $t=1$ there is a pole of order $4d+1$, and the sum in \eqref{specOPE} equals minus its residue. Writing $u=t-1$, this is the coefficient of $u^{4d-1}$ in the expansion of the function
\begin{equation}\label{uexpansion}
n(-1)^{2d+1}\frac{(1+u)^{2d-1}}{(1+u)^n-1}\,.
\end{equation}
Now,  it is clear that this coefficient is zero for $n=1$, since the expansion of $u^{-1}(1+u)^{2d-1}$ has no term of order $u^{4d-1}$. It remains to show that it is a polynomial of degree $4d$. To do this we re-write \eqref{uexpansion} as
\begin{equation}
(-1)^{2d+1}(1+u)^{2d-1}\left(\sum_{k=0}^\infty\frac{n^k}{(k+1)!}(\ln(1+u))^{k+1}\right)^{-1}\,.
\end{equation}
When we expand the sum in large parentheses in powers of $u$, the leading term is $u^1$, and after that the coefficient of $u^m$ is a polynomial in $n$ of degree $m-1$. It follows that, when we expand the whole expression in powers of $u$, the leading term is $u^{-1}$, and after that the coefficient of $u^m$ is a polynomial in $n$ of degree $m+1$. So in particular the coefficient of $u^{4d-1}$ is a polynomial of degree $4d$.

\bibliography{ref}

\providecommand{\href}[2]{#2}\begingroup\raggedright\begin{thebibliography}{10}

\bibitem{PhysRevLett.100.070502}
M.~M. Wolf, F.~Verstraete, M.~B. Hastings, and J.~I. Cirac, {\it Area laws in
  quantum systems: Mutual information and correlations},  {\em Phys. Rev.
  Lett.} {\bf 100} (Feb, 2008) 070502,
  [\href{http://xxx.lanl.gov/abs/0704.3906}{{\tt 0704.3906}}].

\bibitem{Holzhey:1994we}
C.~Holzhey, F.~Larsen, and F.~Wilczek, {\it {Geometric and renormalized entropy
  in conformal field theory}},  {\em Nucl. Phys.} {\bf B424} (1994) 443--467,
  [\href{http://xxx.lanl.gov/abs/hep-th/9403108}{{\tt hep-th/9403108}}].

\bibitem{Calabrese:2009ez}
P.~Calabrese, J.~Cardy, and E.~Tonni, {\it {Entanglement entropy of two
  disjoint intervals in conformal field theory}},  {\em J. Stat. Mech.} {\bf
  0911} (2009) P11001, [\href{http://xxx.lanl.gov/abs/0905.2069}{{\tt
  0905.2069}}].

\bibitem{Ryu:2006bv}
S.~Ryu and T.~Takayanagi, {\it Holographic derivation of entanglement entropy
  from {AdS/CFT}},  {\em Phys. Rev. Lett.} {\bf 96} (2006) 181602,
  [\href{http://xxx.lanl.gov/abs/hep-th/0603001}{{\tt hep-th/0603001}}].

\bibitem{Ryu:2006ef}
S.~Ryu and T.~Takayanagi, {\it Aspects of holographic entanglement entropy},
  {\em JHEP} {\bf 08} (2006) 045,
  [\href{http://xxx.lanl.gov/abs/hep-th/0605073}{{\tt hep-th/0605073}}].

\bibitem{VanRaamsdonk:2009ar}
M.~Van~Raamsdonk, {\it {Comments on quantum gravity and entanglement}},
  \href{http://xxx.lanl.gov/abs/0907.2939}{{\tt 0907.2939}}.

\bibitem{Fursaev:2006ih}
D.~V. Fursaev, {\it {Proof of the holographic formula for entanglement
  entropy}},  {\em JHEP} {\bf 09} (2006) 018,
  [\href{http://xxx.lanl.gov/abs/hep-th/0606184}{{\tt hep-th/0606184}}].

\bibitem{Gross:1998gk}
D.~J. Gross and H.~Ooguri, {\it {Aspects of large N gauge theory dynamics as
  seen by string theory}},  {\em Phys. Rev.} {\bf D58} (1998) 106002,
  [\href{http://xxx.lanl.gov/abs/hep-th/9805129}{{\tt hep-th/9805129}}].

\bibitem{Maldacena:1998bw}
J.~M. Maldacena and A.~Strominger, {\it {AdS(3) black holes and a stringy
  exclusion principle}},  {\em JHEP} {\bf 12} (1998) 005,
  [\href{http://xxx.lanl.gov/abs/hep-th/9804085}{{\tt hep-th/9804085}}].

\bibitem{MR1796805}
M.~A. Nielsen and I.~L. Chuang, {\em Quantum computation and quantum
  information}.
\newblock Cambridge University Press, Cambridge, 2000.

\bibitem{MR2363070}
D.~Petz, {\em Quantum information theory and quantum statistics}.
\newblock Theoretical and Mathematical Physics. Springer-Verlag, Berlin, 2008.

\bibitem{calabrese-2009}
P.~Calabrese and J.~Cardy, {\it Entanglement entropy and conformal field
  theory},  {\em J. Phys. A} {\bf 42} (2009) 504005,
  [\href{http://xxx.lanl.gov/abs/0905.4013}{{\tt 0905.4013}}].

\bibitem{MR0373508}
E.~H. Lieb and M.~B. Ruskai, {\it A fundamental property of quantum-mechanical
  entropy},  {\em Phys. Rev. Lett.} {\bf 30} (1973) 434--436.

\bibitem{MR0345558}
E.~H. Lieb and M.~B. Ruskai, {\it Proof of the strong subadditivity of
  quantum-mechanical entropy},  {\em J. Math. Phys.} {\bf 14} (1973)
  1938--1941. With an appendix by B. Simon.

\bibitem{MR0339901}
J.~Acz{\'e}l, B.~Forte, and C.~T. Ng, {\it Why the {S}hannon and {H}artley
  entropies are `natural'},  {\em Advances in Appl. Probability} {\bf 6} (1974)
  131--146.

\bibitem{MR0434290}
W.~Ochs, {\it A new axiomatic characterization of the von {N}eumann entropy},
  {\em Rep. Math. Phys.} {\bf 8} (1975), no.~1 109--120.

\bibitem{Lunin:2000yv}
O.~Lunin and S.~D. Mathur, {\it {Correlation functions for M(N)/S(N)
  orbifolds}},  {\em Commun. Math. Phys.} {\bf 219} (2001) 399--442,
  [\href{http://xxx.lanl.gov/abs/hep-th/0006196}{{\tt hep-th/0006196}}].

\bibitem{calabrese-2008}
P.~Calabrese and A.~Lefevre, {\it Entanglement spectrum in one-dimensional
  systems},  {\em Phys. Rev. A} {\bf 78} (2008) 032329,
  [\href{http://xxx.lanl.gov/abs/0806.3059}{{\tt 0806.3059}}].

\bibitem{Nishioka:2009un}
T.~Nishioka, S.~Ryu, and T.~Takayanagi, {\it {Holographic Entanglement Entropy:
  An Overview}},  {\em J. Phys.} {\bf A42} (2009) 504008,
  [\href{http://xxx.lanl.gov/abs/0905.0932}{{\tt 0905.0932}}].

\bibitem{Hubeny:2007xt}
V.~E. Hubeny, M.~Rangamani, and T.~Takayanagi, {\it {A covariant holographic
  entanglement entropy proposal}},  {\em JHEP} {\bf 07} (2007) 062,
  [\href{http://xxx.lanl.gov/abs/0705.0016}{{\tt 0705.0016}}].

\bibitem{Wald:1993nt}
R.~M. Wald, {\it {Black hole entropy is the Noether charge}},  {\em Phys. Rev.}
  {\bf D48} (1993) 3427--3431,
  [\href{http://xxx.lanl.gov/abs/gr-qc/9307038}{{\tt gr-qc/9307038}}].

\bibitem{Maldacena:2001kr}
J.~M. Maldacena, {\it {Eternal black holes in Anti-de-Sitter}},  {\em JHEP}
  {\bf 04} (2003) 021, [\href{http://xxx.lanl.gov/abs/hep-th/0106112}{{\tt
  hep-th/0106112}}].

\bibitem{Headrick:2007km}
M.~Headrick and T.~Takayanagi, {\it {A holographic proof of the strong
  subadditivity of entanglement entropy}},  {\em Phys. Rev.} {\bf D76} (2007)
  106013, [\href{http://xxx.lanl.gov/abs/0704.3719}{{\tt 0704.3719}}].

\bibitem{Hubeny:2007re}
V.~E. Hubeny and M.~Rangamani, {\it {Holographic entanglement entropy for
  disconnected regions}},  {\em JHEP} {\bf 03} (2008) 006,
  [\href{http://xxx.lanl.gov/abs/0711.4118}{{\tt 0711.4118}}].

\bibitem{Calabrese:2004eu}
P.~Calabrese and J.~L. Cardy, {\it {Entanglement entropy and quantum field
  theory}},  {\em J. Stat. Mech.} {\bf 0406} (2004) P002,
  [\href{http://xxx.lanl.gov/abs/hep-th/0405152}{{\tt hep-th/0405152}}].

\bibitem{Henningson:1998gx}
M.~Henningson and K.~Skenderis, {\it {The holographic Weyl anomaly}},  {\em
  JHEP} {\bf 07} (1998) 023,
  [\href{http://xxx.lanl.gov/abs/hep-th/9806087}{{\tt hep-th/9806087}}].

\bibitem{Dixon:1986qv}
L.~J. Dixon, D.~Friedan, E.~J. Martinec, and S.~H. Shenker, {\it {The Conformal
  Field Theory of Orbifolds}},  {\em Nucl. Phys.} {\bf B282} (1987) 13--73.

\bibitem{Furukawa:2008uk}
S.~{Furukawa}, V.~{Pasquier}, and J.~{Shiraishi}, {\it {Mutual Information and
  Boson Radius in a c=1 Critical System in One Dimension}},  {\em Phys.\ Rev.\
  Lett.} {\bf 102} (May, 2009) 170602--+,
  [\href{http://xxx.lanl.gov/abs/0809.5113}{{\tt 0809.5113}}].

\bibitem{Dijkgraaf:2000fq}
R.~Dijkgraaf, J.~M. Maldacena, G.~W. Moore, and E.~P. Verlinde, {\it {A black
  hole farey tail}},  \href{http://xxx.lanl.gov/abs/hep-th/0005003}{{\tt
  hep-th/0005003}}.

\bibitem{MinwallaRaju}
S.~Minwalla and S.~Raju. Unpublished work.

\bibitem{Krasnov:2000zq}
K.~Krasnov, {\it {Holography and Riemann surfaces}},  {\em Adv. Theor. Math.
  Phys.} {\bf 4} (2000) 929--979,
  [\href{http://xxx.lanl.gov/abs/hep-th/0005106}{{\tt hep-th/0005106}}].

\bibitem{Witten:2007kt}
E.~Witten, {\it {Three-Dimensional Gravity Revisited}},
  \href{http://xxx.lanl.gov/abs/0706.3359}{{\tt 0706.3359}}.

\bibitem{Yin:2007gv}
X.~Yin, {\it {Partition Functions of Three-Dimensional Pure Gravity}},
  \href{http://xxx.lanl.gov/abs/0710.2129}{{\tt 0710.2129}}.

\bibitem{Yin:2007at}
X.~Yin, {\it {On Non-handlebody Instantons in 3D Gravity}},  {\em JHEP} {\bf
  09} (2008) 120, [\href{http://xxx.lanl.gov/abs/0711.2803}{{\tt 0711.2803}}].

\bibitem{Maloney:2007ud}
A.~Maloney and E.~Witten, {\it {Quantum Gravity Partition Functions in Three
  Dimensions}},  {\em JHEP} {\bf 02} (2010) 029,
  [\href{http://xxx.lanl.gov/abs/0712.0155}{{\tt 0712.0155}}].

\end{thebibliography}\endgroup
\bibliographystyle{JHEP}

\end{document}